*Three-Dimensional Hieratical Twists in Polar Fluids:*
*Chirality Regulation by Ultra-Low Electric Field*

Hiroya Nishikawa*, Dennis Kwaria, Atsuko Nihonyanagi, and Fumito Araoka*
RIKEN Center for Emergent Matter Science, 2-1 Hirosawa, Wako, Saitama 351-0198, Japan
E-mail: hiroya.nishikawa@riken.jp; fumito.araoka@riken.jp

**Abstract**
Recently discovered helical polar fluid adopts a spontaneous chiral symmetry breaking (CSB) driven by polarization escape and conformational chirality. Ferroelectric nematic and smectic phases are intrinsically chiral in the ground state and can be stabilized in an extrinsic twisted configuration through surface anchoring. Herein, we introduce extrinsic CSB as a novel technique in chiral engineering. To demonstrate this concept, we constructed the extrinsic structure of a helielectric conical mesophase (HEC)—three-dimensional chiral system. Considering the challenges of controlling chirality at the macroscopic scale owing to magnetic fields, light, and fluid vortex motion, the proposed three-dimensional chiral system enables chirality (twist) modulation through an ultralow electric field, thereby controlling unique diffraction pattern and circular polarized light-switching capabilities.

## 1. Introduction

Symmetry breaking (SB) is a widespread phenomenon, observed even at the nanoscale. In particular, chiral symmetry breaking (CSB) plays a critical role in the origins of life and the structural evolution of the universe [1,2]. At the nanometer scale, CSB offers promising pathways for constructing functional chiral nanostructures and materials. External perturbations—including magnetic fields[3–5], swirling in fluids[6–11], and circularly polarized (CP) lights or optical vortex [12–16]—can switch chirality in otherwise achiral molecules. However, these methods typically result in barely noticeable enantiomeric excesses, which are insufficient for realizing high-performance nanoscale applications such as CP light sources, optical meta-elements, and quantum devices. Furthermore, achieving inversion of bulk chirality remains a significant challenge, demanding new physical driving forces or innovative approaches to chiral material engineering.

From 2017 to 2020, spontaneous polar SB was observed in the nematic phase. The resulted ferroelectric nematic ($N_F$) phase (**Figure 1d**)[17–20] is characterized by a nondegenerate local nematic director ($\mathbf{n} \neq -\mathbf{n}$). This phase holds fundamental[21–31] and practical[32–40] interests. In 2024, left- and right-handed twisted structures that are formed due to spontaneous CSB in achiral molecules, namely (i) the chiral ground state [30] of the $N_F$ phase (Figure 1b and e) and (ii) helical structures (heliconical nematic[41,42] and heliconical smectic C (Figure 1h)[43]) were discovered. Polarization escape between neighboring polar molecules induces twists that result in intrinsic structures with fully or partially compensated. These are fascinating and diverse structures formed by spontaneous CSB driven by electric dipole interactions in a fluid, in contrast to the spontaneous CSB associated with magnetic dipoles via the Dzyaloshinskii-Moriya interaction in solid materials [44].

Recently, we discovered that the chiral nematic ferroelectric ($N_F$) ground state extends into the ferroelectric smectic phase, where an extrinsic twist structure induced by degenerated surface anchoring is inherited as twisted polar blocks [45]. Building on this finding, we propose utilizing such extrinsic chiral structures to induce three-dimensional (3D) twists in higher-order polar orderings (Figure 1h, i). This unique twist structure does not appear in a synparallel-rubbed (SP) cell (Figure 1a) because of the unnecessity of twist. In contrast, an antiparallel rubbed (AP) cell (Figure 1b) generates enantiomeric $N_F$ domains in a 50:50 ratio. When an electric field (*E*-field) is applied parallel to the rubbing direction (**R** ∥ **E**) (Figure 1c), the twist is unwound and hence there is no chirality (Figure 1f). However, applying a perpendicular field (**R** ⊥ **E**) (Figure 1c) creates a macroscopically observable homochiral domain of the nano-heliconical structure, and reversing the field leads to chiral inversion (Figure 1g). This strategy provides a versatile platform for exploring 3D chiral systems by embedding twist into helielectric phases.

This study examined polar molecules, **1**-*n* (where n = 1–3) (**Figure 2a**), with large dipole moments (ca. 10 Debye) that facilitate polarization escape and chiral conformation. These molecules exhibit an enantiotropic helielectric conical mesophase (HEC), in which a heliconical polar ordering of a few hundred nanometers pitch is formed (Figure 1h), by cooling through the direct $N_F$–HEC phase transition. Furthermore, when **1**-*3* was confined in an AP cell, extrinsic and intrinsic chiral structures were found to coexist (Figure 1i). Remarkably, this chiral state possessed a three-dimensional twist and exhibited two in-plane *E*-field response modes: diffraction switching (Figure 1j) and sign inversion of CP in transmission and reflection (Figure 1k). Both modes were achieved under ultralow *E*-fields (<1.0 V/μm), offering a new approach for controlling CSB.

## 2. Materials

We designed new polar molecules, **1**-*n* (n = 1–3) (**Figure 2a**) based on the existing experimental facts: (1) elongated molecules with structural distortion sites, such as ester or difluoromethylether groups, exhibiting heliconical $N_F$/smectic C ($SmC_F$) phases[41–43], (2) polar molecules with alkyl ester moieties [46] exhibiting $N_F$ and $SmA_F$ phases, and (3) introduction of alkyl ester groups into the LC core induces CSB [47]. The synthesized **1**-*n* (n = 1–3) exhibited pronounced dipole moments ($\mu$ = 9.9–10.2) and relatively large dipolar angles ($\beta$ = 20.0–18.0), corresponding to the angle between the permanent dipole moment and long molecular axis (Figure S14 and Table S1, Supporting Information). The features and the ability to freely rotate around ester/biphenyl units can mitigate polarization loss between neighboring molecules and promote chiral conformations of the molecules (Note S1, Supporting Information). All components of the **1**-*n* series exhibit a uniform pattern of enantiotropic $N_F$–HEC phases (Figure 2b, and Note 2, Supporting Information).

### 3. Phase transition behavior

The DSC curves for samples *1*-**3** (Figures 2c, d, and S15, Table S2, Supporting Information) exhibited three distinct exothermal peaks during heating and cooling. At the phase transition of $T_{N\text{–}IL}$, the domain boundary formed owing to a $2\pi$ twist wall appears in an AP cell [48], which is an extrinsic chiral structure due to the polar coupling to the surface anchoring. The generated *L*- and *R*-handed twisted $N_F$ domains separated by a $2\pi$-wall are observable through uncrossed polarizers (Figure 2e, bottom). A second transition to an unknown phase occurs notably with a tiny exothermal peak (146.9 °C; Figures. 2d and S15, Supporting Information). Coinciding with this phase transition, regularly ordered bright and dark stripes along the rubbing direction emerge under POM. Upon further cooling, each bright stripe gradually modulates like a winding filament (Insets in Figure 2f) (Figure 2f). Furthermore, the winding filaments gradually develop into coarsened coil-like spirals (Figure 2g), expand, and then merge with each other (Figure 2h). Video S1 (Supporting Information) shows the texture evolution from the twisted $N_F$ to the spiral filament states. The $2\pi$-wall between the two twisted domains remains, and the filaments on both sides display opposite winding senses (insets in Figure 2f and g). The domain contrast is inverted by opposite rotation of the analyzer (bottom panels in Figure 2f–h). Thus, the stripe and winding filament states originate from the extrinsic chiral structure of the $\pi$-twist configuration in the $N_F$ phase Figure 2j). When observed orthogonally to the stripe direction, the cell exhibits vivid colors, noticeable to the eye with angle-dependent variation (Figure 2k, l), suggesting a periodic internal structure, most likely local helicity of a few hundred nanometer pitch. Moreover, this colored light scattering is confirmed even in the hemispherical droplet, not in the cell geometry (Figure 2i). Second harmonic (SH) micrography confirmed robust signals in the bright stripe sections (Figure S16, Supporting Information) with using a vertically polarized fundamental light, suggesting polar/helical ordering orienting perpendicular to the stripe direction, consistent with the helielectric (HEC) phase in Figure 1h. However, the dark regions exhibit no SH signal, indicating undetectable polar order along the viewing direction. As the $\pi$-twist from $N_F$ is maintained in HEC, the helical axis of the HEC structure twists along the $t$-direction (Figure 1i). In this situation, the helical axis at the cell mid-plane initially directs in ($h_1$), perpendicular to the rubbing direction. A secondary in-plane helix ($h_2$ direction) is then generated in the midcell, likely to reduce the net electrostatic potential (Figure 2m, n). This secondary helix results the stripe pattern in the rubbing direction. Upon further cooling, the HEC structure undergoes further winding in the $h_1$ direction, resulting in the spiraling along the HEC structure, as shown in Figure 2f–h. In this process, twist senses of the $\pi$-twist and the HEC structure are coupled, as suggested by the correlation between the $\pi$-twist and handedness of the filaments. The reason of this chiral correlation despite the achiral system is yet unclear. However, this is intuitively possible via the elastic coupling, since the helical pitch of the HEC structure should be ~400 nm (red reflection at 151°C, Figure 2i), whereas the $\pi$-twist pitch measures 5 μm, indicating comparable degrees of twist. Hereafter, the HEC phase is classified into $HEC^{HT}$, $HEC^{MT}$, and $HEC^{LT}$ for high-, middle-, and low-temperature states, respectively (Figure 2f–h).

**4. Structural characterization**

**Figure 3a–d** and 3e present the two-dimensional small-angle X-ray diffraction (SAXS) patterns of **1**-*3* molecules under a ~1 T magnetic ($M$-)field and the corresponding 1D X-ray diffractogram, respectively. During the transition from the $N_F$ to the HEC phase, the diffraction pattern rapidly changes from diffuse arcs to distinct split arcs (Figure 3b and c). This bifurcated pattern indicates an SmC bookshelf configuration, wherein the molecular director aligns with the $M$-field and the layer reflections are tilted. Thus, the HEC phase probably exhibits tilted smectic or dense cybotactic lamellar organization. As the temperature decreases in Figure 3f, the lamellar spacing reduces, aligning with the molecular tilt. The tilt angle ($\theta_{tilt}$) derived from the lamellar spacing or the angle ($\beta$) between split spots (Inset in Figure 3f) demonstrates a strong correlation between the methodologies. The results of wide-angle X-ray diffraction (WAXD) patterns are discussed in Note 3 (Supporting Information). Thus, although the SmC-type long-range correlation is not observed in the HEC phase, this phase likely possesses short-range SmC-like ordering. Extended XRD data are shown in Figure S17–S19 (Supporting Information). Support evidence includes (i) an increase in the full width at half maximum (FWHM) in the low-temperature regions (Figure S17, Supporting Information) and (ii) increased diffuse spots compared to the distinct SmC phase.

## 5. Polar behavior

Broadband dielectric spectroscopy (BDS), polarization reversal current (PRC), and second harmonic generation (SHG) were performed to examine the polar properties in **1**-*3*. The $N_F$ phase displayed high dielectric permittivity ($\varepsilon'_{app}$ ~10 k) as well-known [18,20,33,41,42,45,47], decreasing to $\varepsilon'_{app}$ ~2 k in the $HEC^{LT}$ region through the $HEC^{HT/MT}$ region (**Figure 4a**, 4c and Figure S20, S21a, Supporting Information). The dielectric loss ($\varepsilon''$) exhibited a relaxation peak at ~200 Hz, attributed to the polar collective motion in the $N_F$ phase. A comparable peak is observed in $HEC^{HT}$ at a much lower frequency, whereas in $HEC^{LT}$, it is likely owing to the rotational Goldstone mode on the tilt cone. In the $HEC^{HT/MT}$ regimes, this peak shift is emphasized (Figure 4b and Figure S20, S21b, Supporting Information). An additional high-frequency peak in the kHz range within the $HEC^{LT}$ regime indicates a potential soft mode [49]. As expected, the $N_F$ phase shows a distinct current peak at each half-cycle (Figure 4d), whereas $HEC^{HT/MT}$ has fast and slow peaks (Figure 4e). Cooling accentuates the slow peak that predominates while attenuating the fast peak as the temperature decreases. This trend continues in $HEC^{LT}$, where the fast peak becomes nearly invisible, resulting in a strongly asymmetric slow peak that indicates a residual fast component (Figure 4f). Complete data is shown in Figure S22, Supporting Information. Figure 4g estimates the corresponding coercive fields ($E_c$) to the isolated fast (triangles) and slow (circles) peaks as functions of temperature. The $E_c$ of the slow peak exhibits a convex monotonic increase upon cooling. However, that of the fast peak in the $HEC^{HT/MT}$ range demonstrates a concave monotonic increase (Inset in Figure 4g), suggesting a unique polarization reversal mechanism. The gradual temperature dependency of the slow peak indicates bulk polarization reversal, specifically the reorientation of the polar smectic structure. The origin of the fast peak is likely attributed to changes in the molecular tilt within the smectic layer and/or the winding/unwinding of helices. The cumulative $P_s$ in the HEC regions in Figure 4g exceeds 3.5 μC $cm^{-2}$ but is almost constant irrespectively of temperature, unlike the $N_F$ phase exhibiting marked Ps ranging from 1.0 to 3.0 μC $cm^{-2}$ upon cooling. Strong SHG signals are consistently observed across all polar phases, including the crystalline (K) phase, during heating and cooling (Figure 4h), validating macroscopic polar ordering in the HEC phase.

## 6. *E*-field-driven bistable switching of tilted diffraction

As demonstrated, $HEC^{HT/MT}$ exhibits the striped texture in the AP cells. When a stepwise direct current (DC) *E*-field is applied in the rubbing direction (**Figure 5a**), the stripes begin to collapse (Figure 5b, panel [iii]) at +0.3 V $mm^{-1}$, forming a uniform texture at $E$ = +1.0 V $mm^{-1}$ (Figure 5b, panel [v]). Reducing the field to +0.5 V $mm^{-1}$ reinstates the stripes but now tilted by ~20° from their original orientation (Figure 5b, panel [vi, vii], denoted as (+) stripe), which is maintained even after the field is removed. Reversing the field polarity changes the orientation of the stripe tilt (Figure 5b, panels [xi, xii], [−] stripe), indicating that the tilt direction is linked to the *E*-field direction. This operation is reproducible (Video S2, Supporting Information), and the (±) stripe patterns were memorized for over 120 h (Figure S23, Supporting Information). The visible stripes in the optical texture serve as a phase grating plate, allowing blue light to be observed by the naked eye when viewed from above (Figure 5c). Therefore, this diffraction color is also sensitive to the viewing position (Figure S24, Supporting Information) but distinguishable from the reflective color seen in Figure 2k and 2l. Laser diffraction confirmed unique spots corresponding to the diffraction angles predicted from the strip pitch (Insets at upper right corner in Figure 5c–e), with the positions of the diffracted spots varying on the basis of the field polarity (Video S3, Supporting Information). Complete data are shown in Figure S25 and S26, Supporting Information.

**7. Ultralow *E*-field driven chirality switching**

Finally, we conducted a comparable experiment with an *E*-field perpendicular to the rubbing direction (**Figure 6a**). Figure 6b shows the POM texture changes in the $HEC^{HT}$ phase during application of an *E*-field. The POM texture started changing at +0.04 V $mm^{-1}$. At +0.2 V $mm^{-1}$, the original vertical stripes evolved into elongated horizontal stripes with increased pitch. The *E*-field polarity is indicated from top to bottom in the image. Furthermore, increasing the field to +0.5 V $mm^{-1}$ produces a uniform state (*A*[+]-state). When the field was decreased, the back transition restores the vertical stripes almost completely at 0 V $mm^{-1}$. A similar phenomenon was observed in the negative field: at −0.5 V $mm^{-1}$ (*A*[−]-state), the texture viewed through crossed polarizers replicated the positive field. However, circular polarizers demonstrated a disparity in brightness, signifying chirality (Figure 6c–e). Circular dichroism (CD) spectroscopy validates a strong CD signal at $|\mathbf{E}|$= 1 V $mm^{-1}$ (Figure 6f and S27, Supporting Information). The $N_F$ phase at 149°C demonstrates negligible CD, whereas cooling to 148.7°C leads to a prominent CD peak owing to selective reflection in a vertically aligned helical structure. The CD peak exhibited a blue shift, as shown in the inset of Figure 6g, suggesting a decrease in helical pitch. Interestingly, the CD sign reversed between *A*(±)-states with $|\mathbf{E}| < 1$ V $mm^{-1}$ (Figure 6g and Figure S28; Video S4, Supporting Information). In contrast to previous studies[50], our configuration facilitates reversible chirality modulation using ultralow *E*-fields. Complete data are shown in Figure S29 and S30, Supporting Information.

**8. On the possible mechanisms of the *E*-field responses in the HEC phase**

In the HEC phases confined in an AP cell, the $\pi$-twist structure originates from the $N_F$ phase and is associated with the HEC axis, which shows the Bragg reflection owing to the nanoscopic heliconical structure (Figure 2k, l). When a weak $E$-field of ±1.0 V mm$^{-1}$ is applied in the **R ∥ E** geometry, the field compensates the electrostatic potential in the HEC structure and induces unwinding of the secondary helix. This results in a formation of a uniform monodomain of the twisted HEC (Figure 1i), but with the midplane director slightly tilted toward the $E$-field. As the field decreases, the stripes reappear but they are tilted according to the director tilt.

In the **R ⊥ E** geometry, the switching mechanism is complex compared to that in the case of SP cell (Note 4, Supporting Information) but can be discussed as follows: the field realigns the midplane direction, inverting the $\pi$-twist orientation, similarly to the twisted $N_F$ domains [19]. A monodomain with a similar twist is established, which, through elastic coupling, changes the handedness of the HEC structure. Then, by further application of the $E$-field, the secondary helix unwinds similarly to the case of the **R ∥ E** geometry. However, in this case, thickening of bulk polar orientation in the field direction is going to happen, but this causes a strong twist deformation near the surface in the cell normal direction. The combined effect to escape from this vertical twist and removal of the horizontal secondary twist promotes vertical orientation of the bulk HEC structure, accounting for the smooth textures with CP reflection ($A$(±)-states) and the corresponding CD signal. Even in this vertical HEC state, we can confirm slight homogeneous surface orientation remaining in the vicinity of the surface (Figure S31, Supporting Information), consistently with the report by Gorecka et al. in which the similar reorientation phenomenon of HEC was observed using AC $E$-field application however without chirality inversion [50]. The mechanism underlying the field-induced horizontal stripe formation is unclear; nevertheless, it is obvious that these horizontal stripes are generated during the chirality reversal process. Because they don't appear, if the $E$-field with the same sign is applied after the $E$-field application. Since the present HEC is a complex hierarchical system, direct observation or mapping of its 3D structure is needed to fully understand the system. For this sake, the state-of-the-art microscopy techniques, such as high-resolution confocal fluorescent microscopy [45], atomic force microscopy [50], or electron microscopy, might be useful. But currently, any of those are not available in such a high temperature fluid phase. However, although the detailed structure and mechanism are not yet clearly given, the sense of the macroscopic π-twist and the nanoscopic HEC structure can evidently couple with each other, and it is controllable with the ultralow DC voltage and its direction. This unprecedented result must be a fascinating, suggesting the new material-based methodology to control nanoscopic chirality simply by the $E$-field application.

## 9. Conclusion

We investigated a series of polar mesogens, **1**-*n* (n = 1–3), featuring alkyl ester terminal groups. These molecules demonstrated a unique $N_F$–HEC phase sequence with the HEC phase displaying a helielectric structure characterized by a short-range SmC order. In an AP cell, the extrinsic chirality of the $N_F$ state is preserved in the HEC phase, resulting in the coexistence of extrinsic and intrinsic twist structures, thereby creating a triple chiral state in the HEC phase defined by stripes oriented perpendicular to the rubbing direction. Remarkably, applying a small *E*-field ($|\mathbf{E}| < 1$ V mm$^{-1}$) parallel (**R** ∥ **E**) or perpendicular (**R** ⊥ **E**) to the rubbing direction changes optical textures and switches CPL. The cross-rubbing configuration allows intrinsic chirality to interact with extrinsic twist, thereby facilitating the selective creation of an upright homochiral domain with either L- or R-handed helical structures. These effects result from the synergistic interplay of extrinsic and intrinsic chirality, establishing a new paradigm in chiral material design and *E*-field-induced chirality regulation.

### Data availability

The authors declare that the data supporting the findings of this study are available within the paper and its supporting information files. All other information is available from the corresponding authors upon reasonable request.

### Acknowledgements

We are grateful to Dr. H. Masunaga (RIKEN, SPring-8 Center) for supporting X-ray measurement in SPring-8 and Dr. Y. Ishida (RIKEN, CEMS) and Dr. H. Koshino (RIKEN, CSRS) for allowing us to use NANOPIX 3.5m system (Rigaku) and JNM-ECZ500 (500 MHz, JEOL), respectively. We wish to thank Dr. T. Nogawa (RIKEN, CSRS) and Dr. E. Imai (RIKEN, CSRS) for the HRFD-MS measurement. We would like to acknowledge the Hokusai GreatWave Supercomputing Facility (project no. RB230008) at the RIKEN Advanced Center for assistance in computing and communication. This work was partially supported by JSPS KAKENHI (JP22K14594; H.N., JP21H01801, JP23K17341; F.A.), RIKEN Special Postdoctoral Researchers (SPDR) fellowship (H.N.), RIKEN Incentive Research Projects (FY2024: H.N.), JST CREST (JPMJCR17N1; F.A.) and JST SICORP EIG CONCERT-Japan (JPMJSC22C3; F.A.).

### Contributions

H.N. and F.A. conceived the project and designed the experiments. H.N. designed molecules and performed DFT calculation. A.N. synthesized and characterized all compound. H.N. carried out DSC, POM, BDS, PRC, CDS and diffraction studies. D.K. performed SHG and SHM measurements. H.N., F.A. and D.K. took place XRD measurements. H.N. and F.A. analyzed data and discussed the results. H.N. and F.A wrote the manuscript, and all authors approved the final manuscript.

**Corresponding authors**

Correspondence to Hiroya Nishikawa or Fumito Araoka

**Competing interests**

The authors declare no competing interests.

# *Figures*

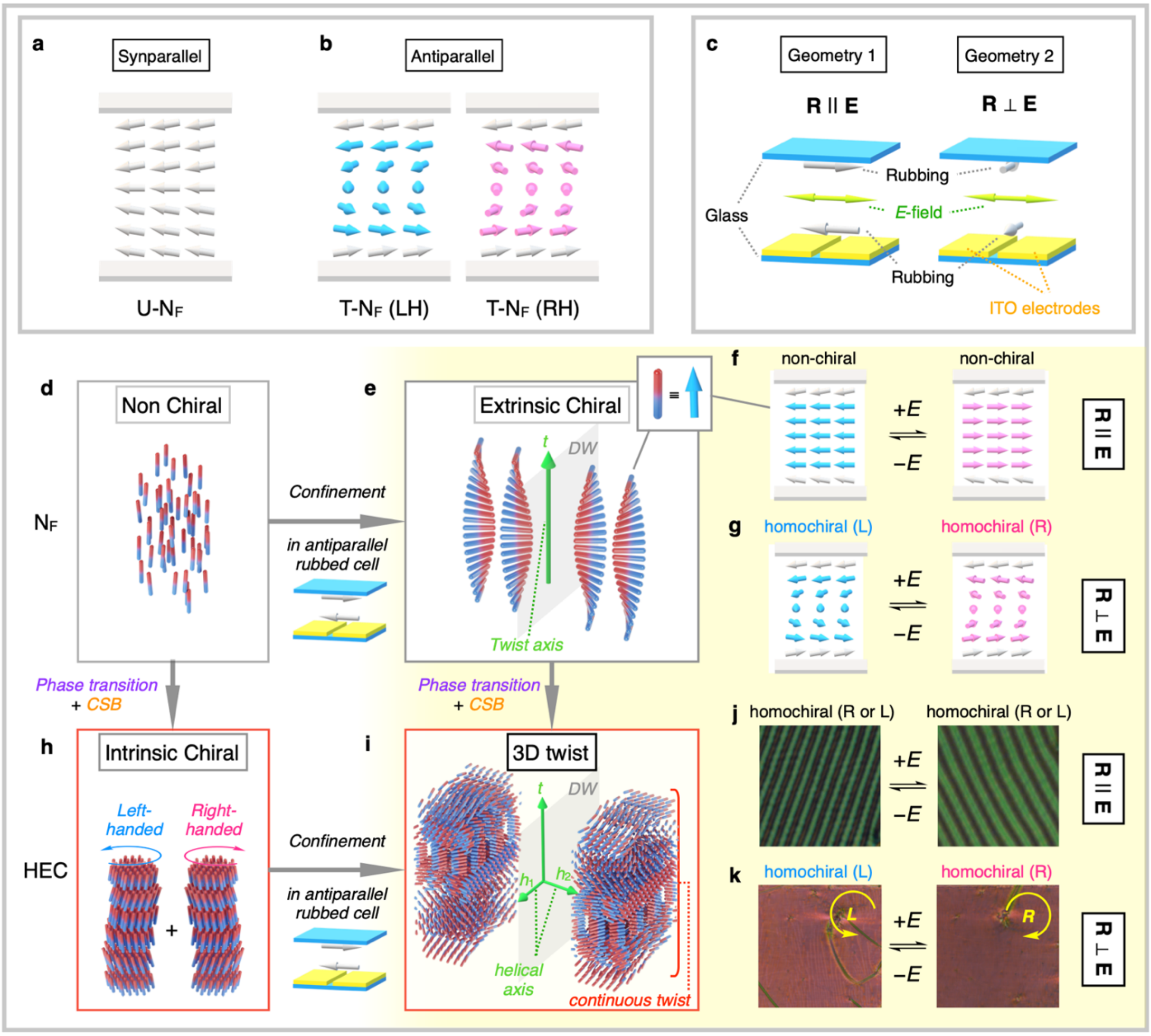

**Figure 1 Extrinsic chiral and triple chiral systems.** Schematic of synparallel (SP: a) and antiparallel (AP: b) rubbed cells and the corresponding director field within the cell. U-$N_F$ and T-$N_F$ represent uniform $N_F$ and twisted $N_F$, respectively. c, Schematic of two types of AP cell geometries: geometry 1 (**R** ‖ **E**) and geometry 2 (**R** ⊥ **E**). **R**: rubbing direction, **E**: electric field direction. Schematic of the $N_F$ (d) and HEC (h) phases in polar fluids. Extrinsic chirality emerges when a polar fluid is confined within an AP cell: one-dimensional (1D) twisted structure of $N_F$ (e). *E*-field response in different geometries, **R** ‖ **E** (f) and **R** ⊥ **E** (g), for the extrinsic chiral state. Note: In the (**R** ‖ **E**) geometry, field reversal induces a polarization reversal by reorientating n and the layer in the twisted $N_F$ states, respectively (f). Thus, the resulting uniform domain is nonchiral. In the (**R** ⊥ **E**) geometry, the 50:50 enantiomeric balance is disrupted, leading to either an *L- or R*-twisted $N_F$ state upon application of an *E*-field (g). The chirality can be inverted by reversing the *E*-field. $N_F$–HEC system: phase transition from $N_F$ (d) to HEC (h). i) In this system, a three-dimensional (3D) twist structure with triple chirality

of HEC is generated in the AP cell. *E*-field response in different geometries, **R** ∥ **E** (j) and **R** ⊥ **E** (k), for the ternary chiral HEC state. Note: In the (**R** ∥ **E**) geometry, the *E*-field influences the formation of a twisted HEC with homochirality (*L* or *R*). During chirality reversal induced by the *E*-field, switching of the tilted diffraction pattern is achieved (j). In the (**R** ⊥ **E**) geometry, a homochiral helix (*L* or *R*) is formed and remains vertical on the substrate upon applying an *E*-field. *E*-field reversal enables chiral inversion, leading to dual CPL switching (k).

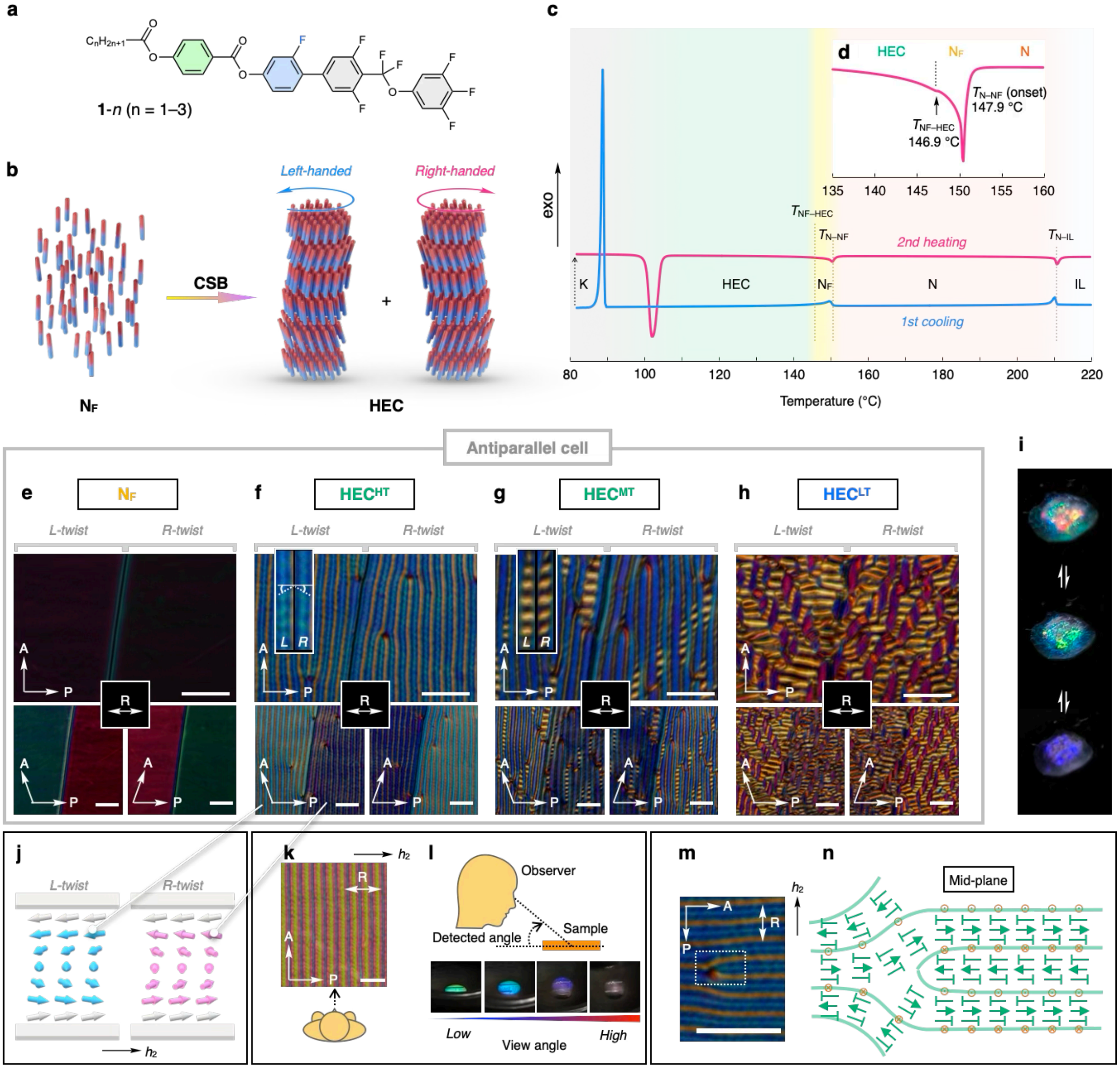

**Figure 2 Phase transition behavior.** a) Chemical structure of **1**-*n* (n = 1–3). b) Schematic of the $N_F$ and HEC phases. DSC curves recorded at heating/cooling rates of 10 (c) and 5 K $min^{-1}$ (d). POM images of the $N_F$ (e) and HEC phases (f–h) in an AP cell with a thickness of 5.0 μm obtained with crossed polarizers (top) and polarizers decrossed by ±20° (bottom). In the $HEC^{HT}$ (f) and $HEC^{MT}$ (g) phases, spiral structures with opposing winding orientations appear prominently on either side of the domain walls. White lines are superimposed in panel (f) to improve the visualization of the spiral configuration. All POM images feature scale bars of 200 μm. i) Macroscopic appearances of LC droplets in the HEC region on the Al substrate in air. Droplets exhibited red, green, and blue colors at 151.0°C, 147.5°C, and 141.0°C, respectively. Note: For the insets in panels (f) and (g), the distance between spiral entities perpendicular the rubbing direction was ~2.0 μm. j) A cross-sectional image of the *L*- and *R*-twist structures in the HEC phase. POM images of the $HEC^{HT}$ state in the AP cell with thicknesses of 10 (k) and 5 μm (m). The detected stripe pattern parallels the rubbing direction (**R**) and secondary helical

axis ($h_2$). l) Detected angle-dependent variation of the reflected color perpendicular to the $h_2$ axis. At large detection angles, the color shifts from yellow green to blue, then to purple, and finally becomes colorless at low detected angles. m) Schematic of the orientation/polarization director in a midplane, including the $h_2$ axis. Note: the existence of fork topologies within the stripe pattern is presented in Figure 2m. The alternating stripes exhibit different birefringence colors, and considering the **n**/**P** directors, it can be assumed that a helical configuration exists along the $h_2$ axis in the midlayer (Figure 2n), with the full pitch being ~10 μm.

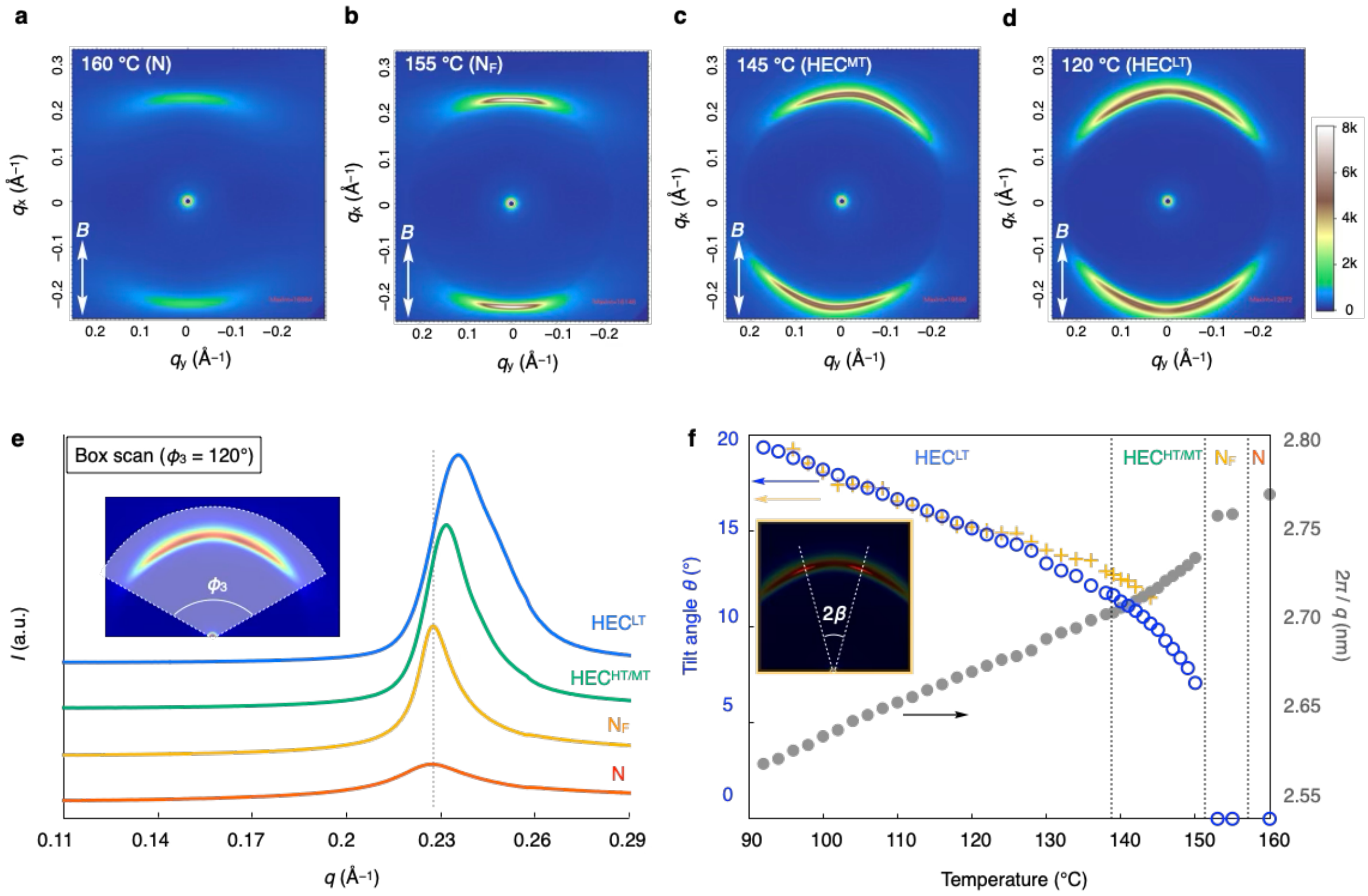

**Figure 3 XRD studies of 1**–*3*. Two-dimensional (2D) SAXS patterns acquired under an *M*-field (~1 T) in the N (a), $N_F$ (b), $HEC^{MT}$ (c), and $HEC^{LT}$ (d) phases. The symbol *B* denotes the direction of the *M*-field. e) 1D X-ray diffractogram along the meridional direction (**n** ∥ **B**) from a box scan area ($\varphi$ = 120°). f) $2\pi/q$ and tilt angle $\theta$ as a function of temperature. Tilt angles obtained from calculations are shown as open circles, while those acquired by image analysis (inset in panel [f]) are indicated by the plus symbol. Notably, because the temperature window of the $HEC^{HT}$ was narrow, accurate identification of the diffraction was difficult. Thus, we investigated X-ray diffraction within the $HEC^{MT}$ and $HEC^{LT}$ regimes.

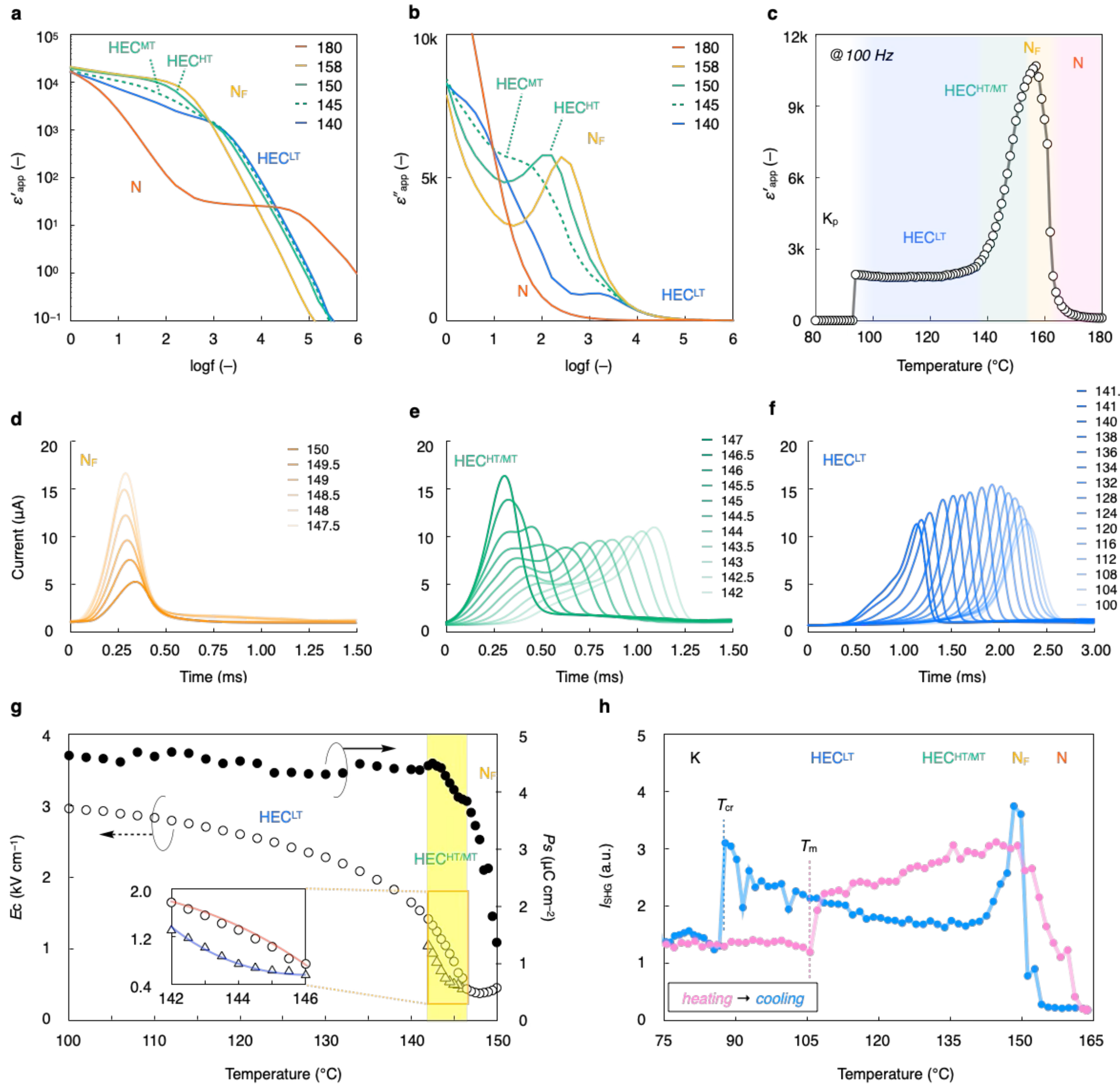

**Figure 4 Polarization behavior of 1**–*3*. a–c) BDS studies in an ITO glass cell without any alignment layer (thickness: 9 µm). The BD spectra are presented as a function of frequency (a), apparent dielectric permittivity ($\varepsilon'_{app}$) (b), and dielectric loss ($\varepsilon''$) as a function of temperature at 100 Hz (c). d–f) PRC profiles recorded in the IPS cell (SP cell, thickness: 5.6 µm, electrode distance: 0.5 mm) using a triangular wave *E*-field ($V_{pp}$ = 800 V, $f$ = 50 Hz). g) Temperature dependence of $E_c$ and polarization density. The inset represents an enlarged profile of $E_c$ vs $T$ within the HEC$^{HT/MT}$ spectrum. h) Temperature dependence of SHG intensity. $T_m$ and $T_{cr}$ represent the melting point and crystallization temperature, respectively.

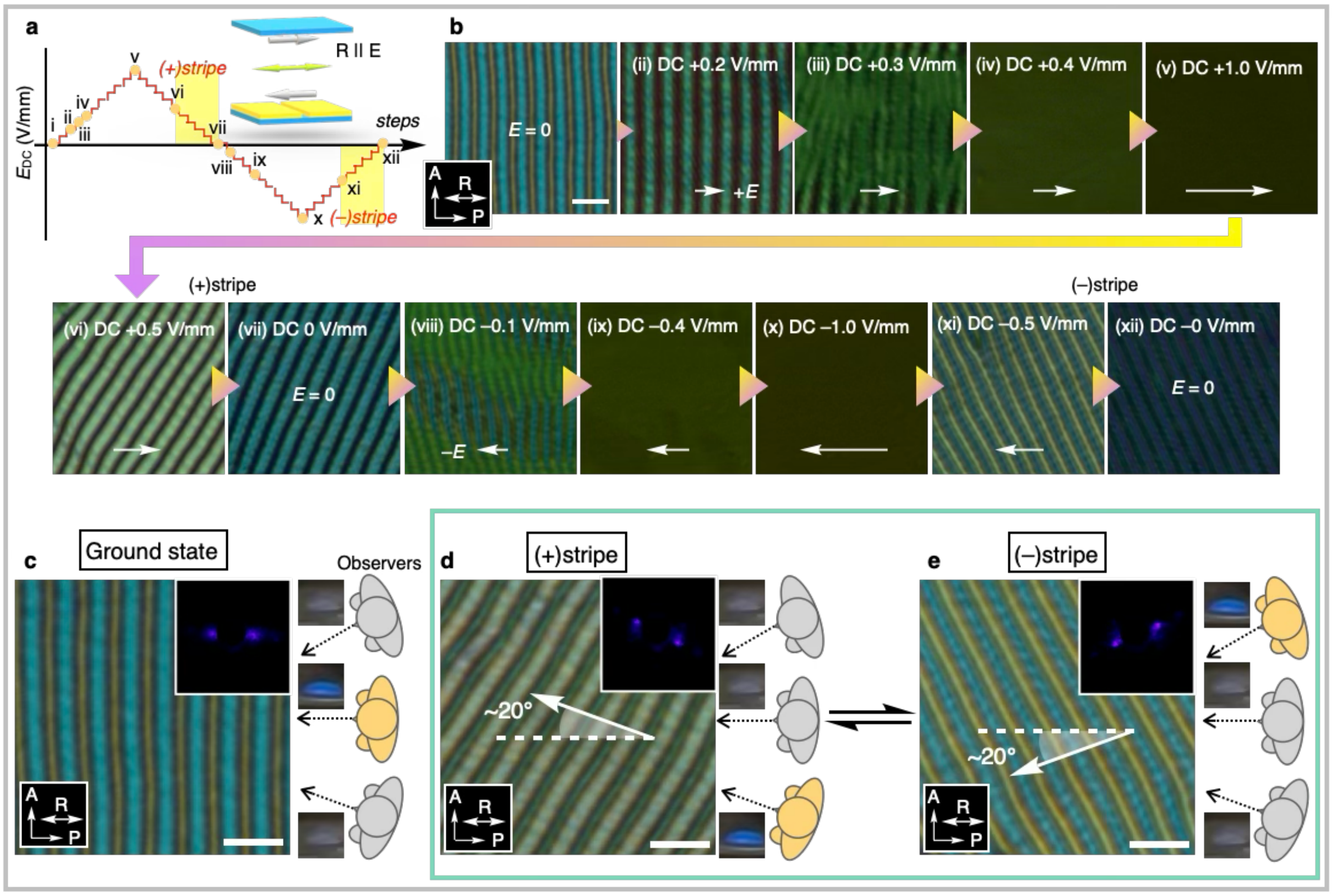

**Figure 5 Dual-grating switching dynamics in the $\text{HEC}^{\text{HT}}$ phase**. a) Schematic of the protocol for the application of a DC *E*-field between +1.0 V/mm and −1.0 V/mm across an AP cell (**R** ∥ **E**), with a thickness of 5 μm and electrode separation of 1 mm. b) POM textural changes in the $\text{HEC}^{\text{LT}}$ phase at 148°C with and without an *E*-field, showing the effect on LC orientation. c–e) Manipulation of the *E*-field in the stripe structure within the $\text{HEC}^{\text{LT}}$ phase at 148°C. The application of a positive and negative *E*-field induced the creation of (+) stripes (d) and (−) stripes (e). Visible and dim color photographs indicate that the reflective color was observable only when viewed parallel to the stripe lines and not observable from the angles (right figures in the corresponding panels). Insets at the upper right corner denote the diffraction patterns corresponding to the stripes in panels (c)–(e) projected onto a screen. All POM images feature scale bars of 10 μm.

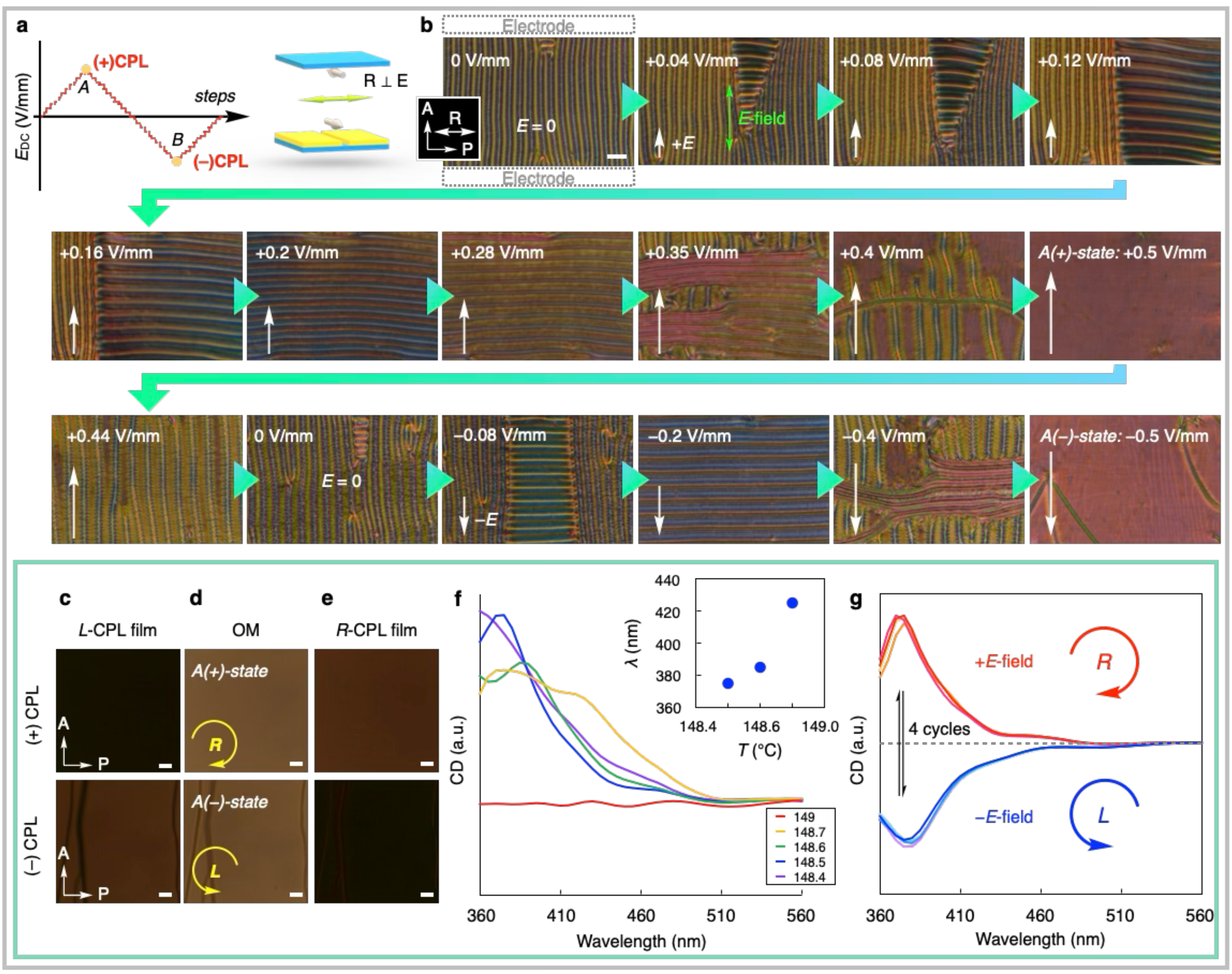

**Figure 6 *E*-field-induced dual CPL switching in the HEC$^{HT}$ phase**. a) Schematic of the protocol for implementing a DC *E*-field ranging from +0.5 V/mm to −0.5 V/mm across an AP cell (**R** ⊥ **E**), with a cell thickness of 10 μm and electrode separation of 1 mm. b) POM textural changes in the HEC$^{HT}$ phase at 148°C with and without an *E*-field. The *A*(+) and *A*(−) states are interchanged by applying *E*-fields of +0.5 V/mm and −0.5 V/mm, respectively. c–e) POM images in the *A*(+) and *A*(−) states (corresponding to panel [b]) under ±*E*-field examined through a nonpolarizing film (d), *L*- (c), or *R*-CPL (e) films. f) CD spectra of *A*(+) and *A*(−) states induced by an *E*-field of ± 1.0 V/mm (four cycles: +*E* → −*E* → +*E* → ⋯). g) Temperature-dependent CD spectra in the *A*(+) state (*E* = + 1.0 V/mm). The inset indicates the central wavelength of the CD peak (*λ*) versus temperature. All POM images feature scale bars of 20 μm.

# Supporting Information

## Methods

### 1. General and materials

**Nuclear magnetic resonance (NMR) spectroscopy**: $^{1}$H, $^{13}$C, and $^{19}$F NMR spectra were recorded on JNM-ECZ500 (JEOL) operating at 500 MHz, 126 MHz, and 471 MHz for $^{1}$H [$^{1}$H{$^{19}$F}], $^{13}$C{$^{1}$H} [$^{13}$C{$^{1}$H,$^{19}$F}] and $^{19}$F [$^{19}$F{$^{1}$H}] NMR, respectively, using the TMS (trimethylsilane) as an internal standard for $^{1}$H NMR and the deuterated solvent for $^{13}$C NMR. The absolute values of the coupling constants are given in Hz, regardless of their signs. Signal multiplicities were abbreviated by s (singlet), d (doublet), t (triplet), q (quartet), quint (quintet), sext (sextet), and dd (double–doublet), respectively.

**High-resolution mass (HRMS) spectroscopy**: The high-resolution field desorption mass spectrometry (HRFD-MS) was performed on AccTOF$^{\mathrm{TM}}$ GCv 4G (JEOL ltd.).

**Density Functional Theory (DFT) Calculation**: Calculations were performed using the Chem3D (pro, 22.2.0.3300) and Gaussian 16 (G16, C.01) softwares (installed at the RIKEN Hokusai GreatWave Supercomputing facility) for MM2 and DFT calculations, respectively. GaussView 6 (6.0.16) software was used to visually analyze the calculation results. [S1] Dipole moments of molecules were calculated using b3lyp-gd3bj/6-311+g(d,p) level. The calculation method is as follows: opt=tight b3lyp/6-311+g(d,p) geom=connectivity empiricaldispersion=gd3bj int=ultrafine.

**Polarized optical microscopy**: Polarized optical microscopy was performed on a polarizing microscope (Eclipse LV100 POL, Nikon) with a hot stage (HSC402, INSTEC) on the rotation stage. Unless otherwise noted, the sample temperature was controlled using the INSTEC temperature controller and a liquid nitrogen cooling system pump (mk2000 and LN2-P/LN2-D2, INSTEC).

**Differential scanning calorimetry (DSC).** Differential scanning calorimetry was performed on a calorimeter (DSC3+, Mettler-Toledo). Rate, 5 and 10 K min$^{-1}$. Cooling/heating profiles were recorded and analyzed using the Mettler-Toledo STARe software system.

**Dielectric spectroscopy.** Dielectric relaxation spectroscopy was performed ranging between 1 Hz and 1 MHz using an impedance/gain-phase analyzer (SI 1260, Solartron Metrology) and a dielectric interface (SI 1296, Solartron Metrology). Prior to starting the measurement of the LC sample, the capacitance of the empty cell was determined as a reference. The temperature was controlled using a homemade heater block and temperature controller. The automatic measurement was took place by the LabVIEW software programming.

**PRC measurement.** PRC measurements were performed in the temperature range of the polar phase under a triangular-wave *E*-field using a ferroelectricity evaluation system (FCE 10, TOYO Corporation), which is composed of an arbitrary waveform generator (2411B), an IV/QV amplifier (model 6252) and a simultaneous A/D USB device (DT9832).

**SHG/SHM measurements.** The SHG/SHM investigation was carried out using a Q-switched DPSS Nd: YAG laser (FQS-400-1-Y-1064, Elforlight) at $\lambda$ = 1064 nm with a 5 ns pulse width (pulse energy: 400 μJ). The primary beam was incident on the LC cell followed by the detection of the SHG signal. The electric field was applied normally to the LC cell. The optical setup is shown in Figure S32.

**Wide- and small-angle X-ray scattering (WAXD, SAXS) analysis.** 2D WAXD measurement was performed using NANOPIX system (Rigaku). 2D SAXS measurement was carried out at BL05B1 in the SPring-8 synchrotron radiation facility (Hyogo, Japan). The samples held in a homemade holder with magnets (Figure S33) was measured at a constant temperature using a temperature controller and a hot stage (mk2000, INSTEC) with high temperature-resistance samarium cobalt magnets (~1 T, Shimonishi Seisakusho Co., Ltd.). The scattering vector $q$ ($q = 4\pi\sin\theta\,\lambda^{-1}$; $2\theta$ and $\lambda$ = scattering angle and wavelength of an incident X-ray beam [1.54 Å (for WAXS) and 1.0 Å (for SAXS)], respectively) and position of an incident X-ray beam on the detector were calibrated using several orders of layer diffractions from silver behenate ($d$ = 58.380 Å). The sample-to-detector distances were 76.64 mm (for WAXS) and 1886.42 mm (for SAXS), where acquired scattering 2D images were integrated along the Debye–Scherrer ring by using software (Igor Pro with Nika-plugin), affording the corresponding one-dimensional profiles.

**Diffraction measurement.** The image of diffraction pattern was captured using a camera. The diffraction pattern from a LC sample (**1**-*3* in the antiparallel rubbed IPS cell) was projected on a screen through a DPSS laser ($\lambda$ = 405 nm, CivilLaser). The measurement setup is shown in Figure S34.

**CD spectra measurement.** CD spectra were recorded using a circular dichroism spectrometer (J-1500, JASCO). A LC sample (**1**-*3* in the antiparallel rubbed IPS cell) was put on the hot stage and then the stage was put into the CD spectrometer. To penetrate an initial beam at measurement area between electrodes, we used an anodized Al pate with a hole (ca. 1mm diameter) (Figure S35). Absorption of the LC film was calculated by subtraction between that of recorded data (LC film + pinhole) and of a pinhole.

**Information of used liquid crystalline (LC) cells**:

*ITO glass cell (EHC)*:

- ITO-coated type, electrode area: 5 × 10 mm
- Experiments: POM (thickness: 10.0 μm) and DR (thickness: 9.0 μm) studies

*Antiparallel-rubbed cell (EHC)*:

- PI-coated type, thickness: 5.0 μm
- Alignment layer: LX-1400
- Experiments: POM studies

*IPS cell (homemade)*:

- PI-coated type, electrode distance: 500 μm, electrode length: 12 mm, thickness: 5.6 μm
- Alignment layer: AL1254
- Rubbing condition: synparallel (**R** ∥ **E**)
- Experiments: PRC studies

*IPS cell (EHC)*:

- PI-coated type, electrode distance: 1 mm, electrode length: 18 mm, thickness: 10.0 μm
- Alignment layer: LX-1400
- Rubbing condition: antiparallel (**R** ⊥ **E**)
- Experiments: POM studies and CD spectra studies

*IPS cell (EHC)*:

- PI-coated type, electrode distance: 500 μm, electrode length: 18 mm, thickness: 5.0 μm
- Alignment layer: LX-1400
- Rubbing condition: antiparallel (**R** ∥ **E**)
- Experiments: POM studies and diffraction experiments

*IPS cell (EHC)*:

- PI-coated type, electrode distance: 1 mm, electrode length: 18 mm, thickness: 10.0 μm
- Alignment layer: LX-1400
- Rubbing condition: synparallel (**R** ∥ **E**)
- Experiments: POM studies

## 2. Synthesis of 1-*3*.

### 2.1. Synthetic route

**1**-*n* (n = 1–3) used in this paper were synthesized by following pathway (**Scheme S1**).

**1-*1*:** R = $CH_3$
**1-*2*:** R = $CH_3CH_2$
**1-*3*:** R = $CH_3(CH_2)_2$

**Scheme S1** Synthetic pathway of **1**-*n* (n = 1–3). a) $B_2pin_2$, Pd(dppf)$Cl_2$- $CH_2Cl_2$, KOAc, 1,4-dioxane, MM400 (30 Hz), 110 °C, 10 min, b) Pd$(OAc)_2$, SPhos, $K_2CO_3$, THF/$H_2O$, 55 °C, 3 h, c) 3,4-dihydro-2*H*-pyran, $Et_2O$, 35 °C, 2.5 h; r.t., 20 h, d) **4**, EDAC-HCl, DMAP, $CH_2Cl_2$, 0 °C, 1 h; r.t., 1 h, e) *p*-TsOH, $CH_2Cl_2$/MeOH, 50 °C, 1.5 h, f) TEA, $CH_2Cl_2$, 0 °C.

**2.2.1.** ***Synthesis of 2-(4-(difluoro(3,4,5-trifluorophenoxy)methyl)-3,5-difluorophenyl)-4,4,5,5-tetramethyl-1,3,2-dioxaborolane*** **(3)**

The synthetic procedure of **3** was reported in previous our works.[S2]

**2.2.2.** ***Synthesis of 4'-(difluoro(3,4,5-trifluorophenoxy) methyl)-2,3',5'-trifluoro-[1,1'-biphenyl]-4-ol*** **(4)**

To a solution of THF (50 mL) and 2M $K_2CO_3$ solution (50 mL) were added **1** (5.23 g, 12.0 mmol), 4-bromo-3-fluorophenol (2.29 g, 12.0 mmol), $Pd(OAc)_2$ (135 mg, 0.60 mmol), and SPhos ligand (493 mg, 1.20 mmol). The reaction mixture was stirred at 55 °C under Ar atmosphere. After 3 hours stirring, the reaction mixture was diluted with $H_2O$ and extracted with AcOEt. After the organic layer was dried over anhydrous $Na_2SO_4$ and evaporated, the residue was purified by flash column chromatography on silica gel (AcOEt/*n*-hexane, 10/90–40/60) to afford the target compound in 94.6% yield (4.77 g, 11.4 mmol) as an orange solid.

$^1H\{^{19}F\}$-NMR (500 MHz, $CDCl_3$): δ 7.31 (d, *J* = 8.6 Hz, 1H), 7.17 (s, 2H), 6.99 (d, *J* = 5.7 Hz, 2H), 6.74-6.70 (m, 2H), 5.48 (br s, 1H)
$^{19}F\{^1H\}$-NMR (471 MHz, $CDCl_3$): δ −61.5, −61.6, −61.6, −110.8, −110.9, −111.0, −114.4, −132.4, −132.4, −163.1, −163.1, −163.2

**2.2.3.** ***Synthesis of 4-((tetrahydro-2H-pyran-2-yl) oxy) benzoic acid*** **(6)**

To a solution of 4-hydroxybenzoic acid (2.76 g, 20.0 mmol) and *p*-TsOH-$H_2O$ (76.1 mg, 0.4 mmol) in $Et_2O$ (60 mL) was added 3,4-dihydro-2*H*-pyran (7.61 mL, 100 mmol) under Ar atmosphere, and the resulting mixture was stirred for 2.5 hours at 35 °C and 20 hours at room temperature. The reaction mixture was evaporated, and to the resulting oil was added *n*-hexane to form a precipitate. The precipitate was filtered off by Kiriyama funnel and dried in vacuo to afford the target compound in 71.2 % yield (3.16 g, 14.2 mmol) as a white powder.

$^{1}$H{$^{19}$F}-NMR (500 MHz, $CD_3OD$): δ 7.97-7.94 (m, 2H), 7.10-7.07 (m, 2H), 5.54 (t, $J$ = 3.3 Hz, 1H), 3.87-3.82 (m, 1H), 3.64-3.60 (m, 1H), 2.05-1.97 (m, 1H), 1.92-1.81 (m, 2H), 1.74-1.66 (m, 2H), 1.62-1.57 (m, 1H)

**2.2.1.** ***Synthesis of 4'-(difluoro(3,4,5-trifluorophenoxy)methyl)-2,3',5'-trifluoro-[1,1'-biphenyl]-4-yl 4-((tetrahydro-2H-pyran-2-yl)oxy)benzoate*** **(7)**

To a solution of 4 (1.99 g, 9.0 mmol) and 2 (3.42 g, 8.13 mmol) in $CH_2Cl_2$ (20 mL) were added WSC (2.34 g, 12.2 mmol) and DMAP (29.8 mg, 0.24 mmol) at 0 °C. After 1 hour stirring at 0 °C and 1 hour stirring at room temperature under Ar atmosphere, the reaction mixture was diluted with $H_2O$ and extracted with $CH_2Cl_2$. After the organic layer was dried over anhydrous $Na_2SO_4$ and evaporated, the residue was purified by flash column chromatography on silica gel ($CHCl_3$/*n*-hexane, 30/70–100/0) to afford the target compound in 73.5% yield (3.48 g, 5.97 mmol) as a white solid.

$^{1}$H{$^{19}$F}-NMR (500 MHz, $CDCl_3$): δ 8.15 (d, $J$ = 8.7 Hz, 2H), 7.48 (d, $J$ = 8.0 Hz, 1H), 7.22 (s, 2H), 7.17-7.15 (m, 4H), 7.00 (d, $J$ = 5.6 Hz, 2H), 5.57 (t, J = 2.8 Hz, 1H), 3.90-3.85 (m, 1H), 3.67-3.63 (m, 1H), 2.05-2.01 (m, 1H), 1.93-1.90 (m, 2H), 1.75-1.68 (m, 2H), 1.65-1.60 (m, 1H)

$^{19}$F{$^{1}$H}-NMR (471 MHz, $CDCl_3$): δ −61.6, −61.7, −61.7, −110.3, −110.4, −110.4, −113.9, −132.3, −132.4, −163.0, −163.0, −163.1

**2.2.1.** ***Synthesis of 4'-(difluoro(3,4,5-trifluorophenoxy)methyl)-2,3',5'-trifluoro-[1,1'-biphenyl]-4-yl 4-hydroxybenzoate*** **(8)**

To a solution of **5** (3.90 g, 6.70 mmol) in $CH_2Cl_2$/MeOH (15/15 mL) was added *p*-TsOH-$H_2O$ (63.6 mg, 0.34 mmol) under Ar atmosphere, and the resulting mixture was stirred 1.5 hours at 50 °C. the reaction mixture was diluted with $H_2O$ and extracted with AcOEt. The organic layer was dried over anhydrous $Na_2SO_4$, concentrated, and dried in vacuo to afford the target compound in 93.8 % yield (3.39 g, 6.27 mmol) as a white solid.

$^{1}$H{$^{19}$F}-NMR (500 MHz, $CDCl_3$): δ 8.14-8.11 (m, 2H), 7.48 (d, $J$ = 8.2 Hz, 1H), 7.22 (s, 2H), 7.17-7.15 (m, 2H), 7.00 (d, $J$ = 5.7 Hz, 2H), 6.95 (m, 2H)

$^{19}F\{^1H\}$-NMR (471 MHz, $CDCl_3$): δ −61.6, −61.7, −61.7, −110.3, −110.3, −110.4, −113.9, −132.3, −132.4, −163.0, −163.0, −163.0

### **2.2.1.** ***Synthesis of 4'-(difluoro(3,4,5-trifluorophenoxy)methyl)-2,3',5'-trifluoro-[1,1'-biphenyl]-4-yl 4-acetoxybenzoate (*****1*****-*****1*****)***

To a solution of 6 (243 mg, 0.45 mmol) in $CH_2Cl_2$ (5 mL) were added acetic anhydride (43.0 μL, 0.45 mmol), $Et_3N$ (75.0 μL, 0.54 mmol) and DMAP (5.5 mg, 0.05 mmol) at room temperature. After 1.5 hours stirring under Ar atmosphere, the reaction mixture was quenched with $H_2O$ and extracted with $CH_2Cl_2$. After the organic layer was dried over anhydrous $Na_2SO_4$ and evaporated, the residue was purified by flash column chromatography on silica gel ($CH_2Cl_2$/*n*-hexane, 20/80–50/50, and 5% MeOH/95% $CH_2Cl_2$) to afford the target compound in 96.2% yield (252 mg, 0.43mmol) as a white solid.

$^1H\{^{19}F\}$-NMR (500 MHz, $CDCl_3$): δ 8.26-8.23 (m, 2H), 7.50 (d, $J$ = 7.9 Hz, 1H), 7.30-7.27 (m, 2H), 7.23 (s, 2H), 7.18-7.16 (m, 2H), 7.00 (d, $J$ = 5.6 Hz, 2H), 2.36 (s, 3H)

$^{19}F\{^1H\}$-NMR (471 MHz, $CDCl_3$): δ −61.7 (t, $J$ = 27.6 Hz), −110.3 (t, $J$ = 27.6 Hz), −113.7 (s), −132.3 (d, $J$ = 22.1 Hz), −163.0 (t, $J$ = 21.9 Hz)

$^{13}C\{^1H\}$-NMR (126 MHz, $CDCl_3$): δ 168.7, 163.8, 159.9, 159.5, 155.2, 152.3, 151.0, 144.6, 140.7, 138.5, 131.9, 130.6, 126.2, 123.3, 122.0 (t, $J$ = 261 Hz), 120.1, 118.4, 113.1, 110.9, 108.9 (t, $J$ = 32.6 Hz), 107.5, 21.1

HRFD-MS ($m/z$, $M^+$) calc. for 582.0713; found, 582.0712

### ***4'-(difluoro(3,4,5-trifluorophenoxy)methyl)-2,3',5'-trifluoro-[1,1'-biphenyl]-4-yl 4-(propionyloxy)benzoated (*****1*****-*****2*****)***

To a solution of 6 (253 mg, 0.47 mmol) in $CH_2Cl_2$ (5 mL) were added propionyl chloride (45.0 μL, 0.52 mmol) and $Et_3N$ (78.0 μL, 0.56 mmol) at 0 °C. After 2 hours stirring under Ar atmosphere, the reaction mixture was quenched with $H_2O$ and extracted with $CH_2Cl_2$. After the organic layer was dried over anhydrous $Na_2SO_4$ and evaporated, the residue was purified by flash column chromatography on silica gel ($CH_2Cl_2$/*n*-hexane, 50/50–90/10) to afford the target compound in 87.4 % yield (244 mg, 0.41mmol) as a white solid.

$^{1}$H{$^{19}$F}-NMR (500 MHz, $CDCl_3$): 8.25-8.23 (m, 2H), 7.51-7.49 (m, 1H), 7.29-7.27 (m, 2H), 7.23 (s, 2H), 7.18-7.16 (m, 2H), 7.00 (d, $J$ = 5.6 Hz, 2H), 2.65 (q, $J$ = 7.5 Hz, 2H), 1.30 (t, $J$ = 7.5 Hz, 3H)

$^{19}$F{$^{1}$H}NMR (471 MHz, $CDCl_3$): δ −61.7 (t, $J$ = 25.7 Hz), −110.3 (t, $J$ = 25.7 Hz), −113.7 (s), −132.3 (t, $J$ = 18.4 Hz), −163.0 (t, $J$ = 20.0 Hz)

$^{13}$C{$^{1}$H}-NMR (126 MHz, $CDCl_3$): δ 172.3, 163.8, 160.0, 159.6, 155.4, 152.3, 151.0, 144.6, 140.8, 138.5, 131.9, 130.6, 126.1, 123.3, 122.1, 120.1 122.0 (t, $J$ = 262 Hz), 118.5, 113.1, 110.9, 108.9 (t, $J$ = 31.5 Hz), 107.5, 27.8, 9.0

HRFD-MS (*m*/*z*, $M^+$) calc. for 596.0870; found, 596.0820

### 2.2.1. *Synthesis of 4'-(difluoro(3,4,5-trifluorophenoxy)methyl)-2,3',5'-trifluoro-[1,1'-biphenyl]-4-yl 4-(butyryloxy)benzoate (1-3)*

To a solution of 3 (104 mg, 0.5 mmol) and 2 (210 mg, 0.5 mmol) in $CH_2Cl_2$ (3 mL) were added WSC (115 mg, 0.6 mmol) and DMAP (6.1 mg, 0.05 mmol) at 0 °C. After 16 hours stirring at room temperature under Ar atmosphere, the reaction mixture was concentrated and purified by flash column chromatography on silica gel ($CH_2Cl_2$/*n*-hexane, 25/75–100/0) to afford the target compound in 33.2% yield (100 mg, 0.166 mmol) as a white solid.

$^{1}$H{$^{19}$F}-NMR (500 MHz, $CDCl_3$): δ 8.25-8.22 (m, 2H), 7.50 (d, $J$ = 8.0 Hz, 1H), 7.28 (d, $J$ = 9.0 Hz, 2H), 7.23 (s, 2H), 7.18-7.16 (m, 2H), 7.00 (d, $J$ = 5.6 Hz, 2H), 2.60 (t, $J$ = 7.3 Hz, 2H), 1.86-1.78 (m, 2H), 1.07 (t, $J$ = 7.4 Hz, 3H)

$^{19}$F{$^{1}$H}-NMR (471 MHz, $CDCl_3$): δ −61.7 (t, $J$ = 25.7 Hz), −110.3 (t, $J$ = 25.7 Hz), −113.7 (s), −132.3 (d, $J$ =18.4 Hz), −163.0 (t, $J$ = 20.0 Hz)

$^{13}$C{$^{1}$H}-NMR (126 MHz, $CDCl_3$): δ 171.4, 163.8, 159.9, 159.5, 155.3, 152.3, 151.0, 144.6, 140.7, 138.5, 131.9, 130.6, 126.1, 123.2, 122.1, 120.1 (t, $J$ = 264 Hz), 118.4, 113.1, 110.9, 107.5, 108.9 (t, $J$ = 31.4 Hz), 36.2, 18.3, 13.6

HRFD-MS (*m*/*z*, $M^+$) calc. for 610.1026; found, 610.1026

**Supporting Notes (Notes S1–S4)**

**Supporting Note 1 | Molecular design.**

In a previous study, Jeong et al. investigated the spontaneous formation and resolution of enantiomers (racemization) in various achiral rod-shaped 4-arylbenzoate esters and demonstrated that CSB originates from axial chirality generated within the biphenyl mesogen connected to the ester bond. This indicates that axial chirality could be a catalyst behind the helielectric phase observed in the archetypal molecules. The molecular structures investigated by Jeong et al. were simple, which featured ester groups coplanar with the adjacent aromatic ring. However, density functional theory (DFT) calculations for **1**–*3* indicated that the ester and aromatic rings exhibit a noncoplanar configuration, suggesting that along with the axial chirality in the biphenyl unit, the molecular conformation generated from rotation around the C–O link of the ester groups may serve as an additional component affecting the helielectric phase. Figure S1 depicts the energy profile for different conformations created by rotating around the C–O link for the two types of benzoate ester (EST1 and EST2) in **1**–*3*. EST1 and EST2 each displayed four energy minima corresponding to the conformations. The potential energy barrier for enantiomers was $\Delta E$ (EST1) = 4.80 × $10^{-24}$ kJ and $\Delta E$ (EST2) = 5.17 × $10^{-24}$ kJ, which was larger than the thermal energy ($k_B$T) at room temperature (4.12 × $10^{-24}$ kJ). The values of $k_B$T in the LC phases were slightly greater than $\Delta E$ (EST1) and $\Delta E$ (EST2) but showed no notable deviation, indicating that C–O rotation of the benzoate esters is crucial in the enantiomeric separation process in bulk material. Furthermore, the permanent dipole moment ($\mu$) and polarization angle ($\beta$) values calculated using the DFT method are summarized in Table S1.

**Supporting Note 2 | Effect of alkyl chain length for 1-*n* homologues.**

The impact of alkyl chain length on HEC phase formation was examined by DSC, POM, and BDS analyses (**1**-*n, n* = 1–3) and clearly confirmed that all homologs exhibited enantiotropic N, $N_F$, and HEC phases (Table S2 and Figure S2–S6). The $HEC^{LT}$ state was not observed in **1**-*1* owing to the narrow temperature window of the HEC phase. Extending the alkyl chain destabilized the $N_F$ phase while stabilizing the smectic phase. This trend is similar to that observed in archetypal molecules such as the **MUT_JK10n** (n = 1–3) series, which exhibit a heliconical $N_F$ phase and is reflected in the thermodynamic stability of $N_F$ and $SmC_F$ phases, indicating that extended alkyl chains facilitate the formation of a lamellar structure, predominantly owing to the remarkable van der Waals interaction between the chains. Notably, all homologs of **1**-*n* show the enantiotropic HEC phase. Therefore, elongating the alkyl chain enhances smectogenicity and provides a direct approach to designing a stable HEC phase.

**Supporting Note 3 | WAXD studies.**

Figure S7a–d depicts 2D WAXD patterns for **1**–*3* under an *M*-field of ~1 T. The 1D diffraction patterns generated from these 2D images using box scan analysis within the $\varphi$ range ($\varphi_1$ = 110° and $\varphi_2$ = 100°) (Figure S7f) are shown in Figure S7e, which reveals that all mesophases have multiple diffraction peaks ($p_1$–$p_6$) spanning small- to wide-angle regions in the meridional direction $\varphi_1$. The principal peak $p_1$ corresponds to the molecular length ($L$ = 2.74 nm), and the ratio of the peak wavenumbers for $p_1$–$p_5$ is 1：2：3：4：5, respectively (Figure S8). However, $p_6$ deviates from this trend. In general, the presence of these higher-order diffraction peaks indicates lamellar structures. Thus, the observation of multiple peaks in the N and $N_F$ phases is attributed to the large correlation length of the smectic-like cybotactic cluster distributed within the nematic ordering. However, such clusters in the N phase above the $N_F$ phase are unusual. A large-diffused halo is detected along the equatorial direction $\varphi_2$, with its 1D profile shown in Figure S7g. The calculated average spacing is ~0.44 nm, which indicates π–π stacking interactions. As the temperature decreases, the spacing and FWHM ($\Delta q_{FWHM}$) decreases (Figure S7h and i), indicating enhanced order. The extended data are presented in Figure S9 and S10.

**Supporting Note 4 | *E*-field switching in synparallel cell (R ⊥ E)**

In the HEC phase confined in SP cells, no stripe structure is formed, and instead, uniform domains are observed. In the field switching of the **R ⊥ E** geometry, near the substrate interface, director cannot reorient due to strong anchoring, whereas the director in the bulk reorients in response to the *E*-field (Figure S11 and S12). Considering the *E*-field gradient in the normal direction within the IPS cell, the director field in the cell becomes as shown in Figure S13, and a structure with chirality is formed. The SHG intensity along the rubbing direction is maximum under zero *E*-field, becoming minimum after *E*-field application. In this case, the SHG intensity along the *E*-field direction reaches its maximum value, confirming the validity of the proposed model. Although the switching mechanism is simple in the case of SP cells (**R ⊥ E**), it is more complex for in AP cells (**R ⊥ E**). The decisive different thing between the AP and SP cells is that the helical axis of the HEC structure stands upright relative to the substrate due to screening based on charge accumulation. However, it remains unclear why effective screening occurs only in AP cells, and this is a topic for future research.

## Supporting Figures (Figures S1–S35)

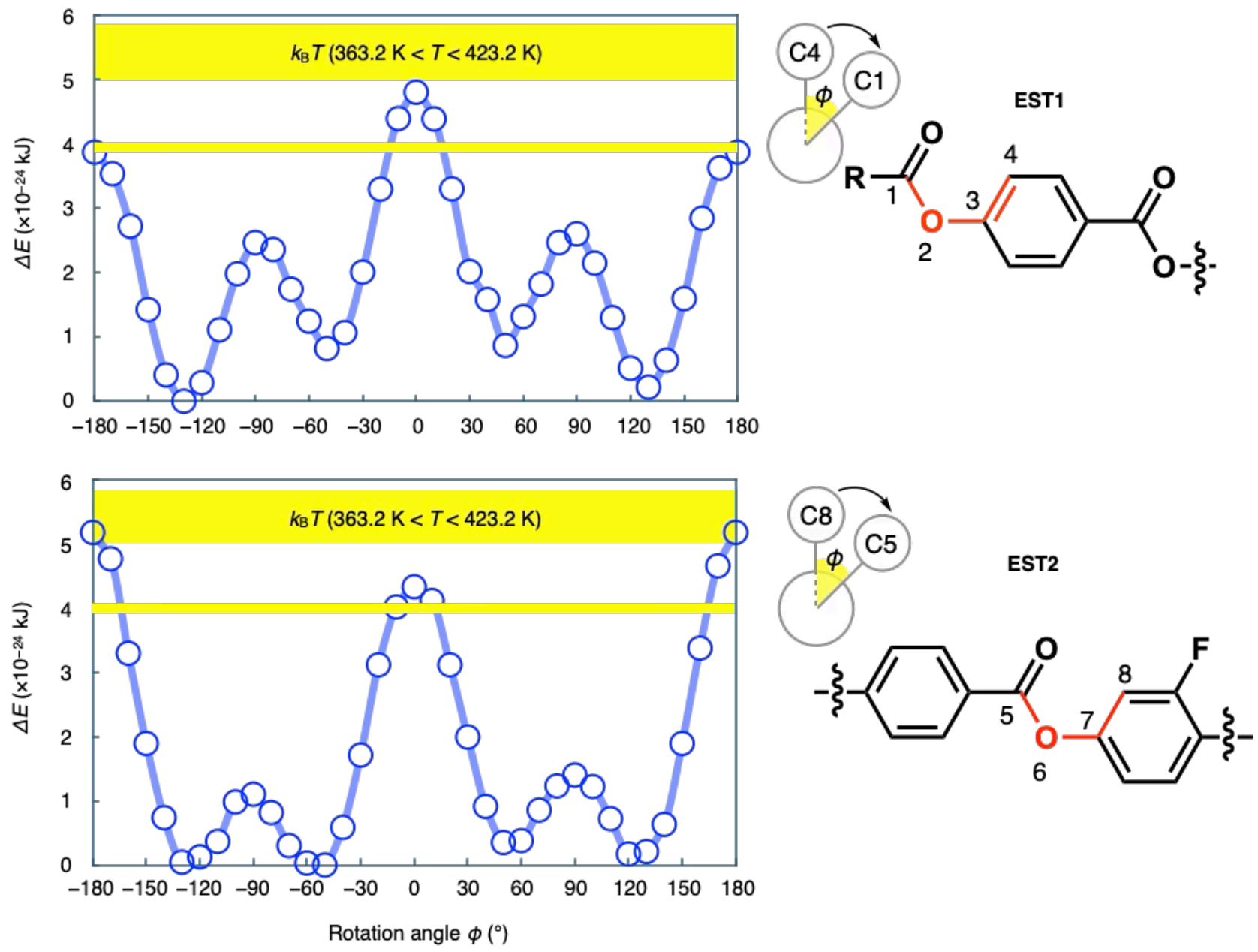

**Figure S1 |** Calculated potential energies of EST1 (a) and EST2 (b) in **1**-*3* as a function of dihedral angle at C(1)–O(2)–C(3)–C(4) and C(5)–O(6)–C(7)–C(8), respectively.

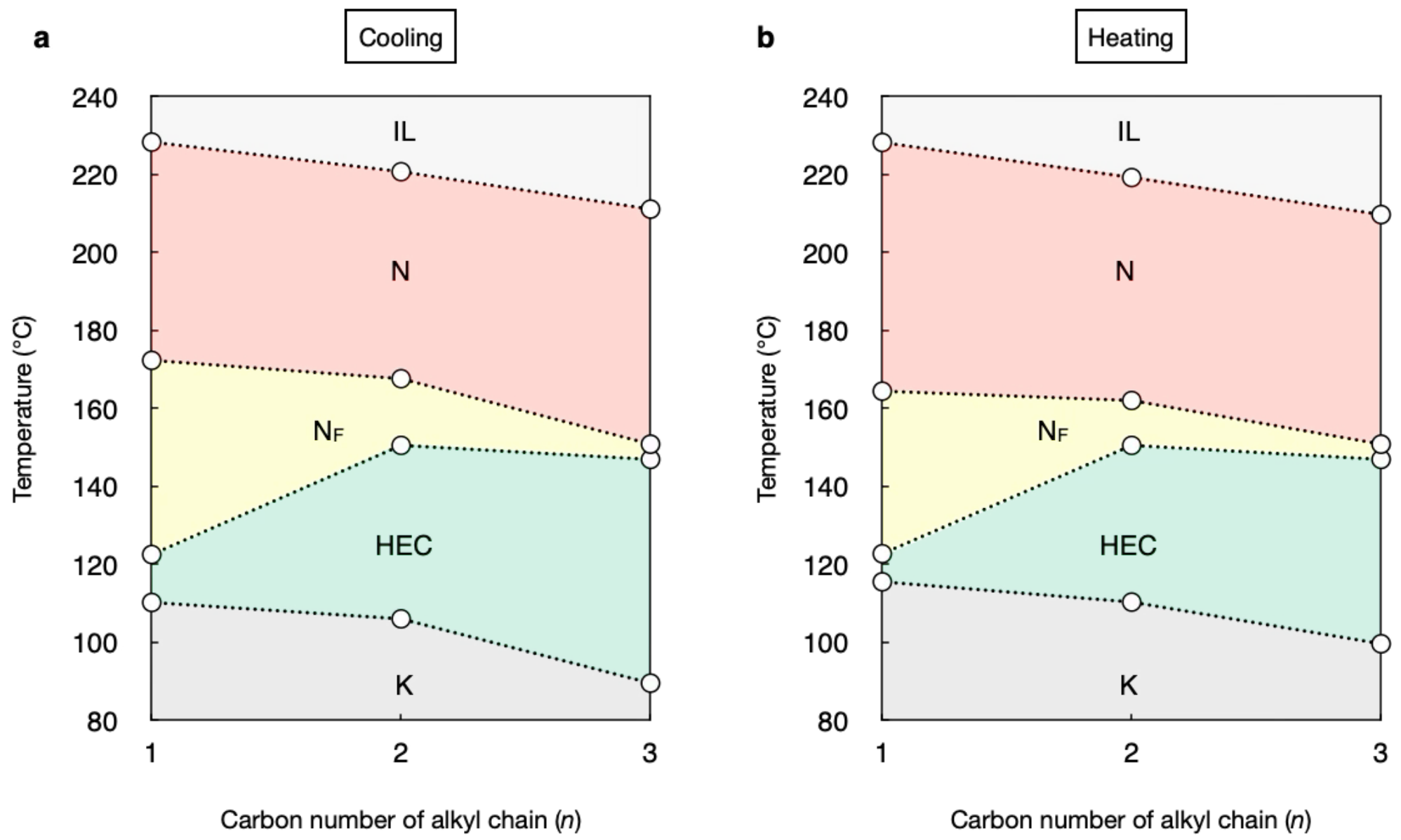

**Figure S2 |** Phase diagram for **1**-*n* (n = 1–3) created using DSC data on cooling (a) and heating (b) process. Abbrev.: IL = isotropic liquid, N = nematic, $N_F$ = ferroelectric nematic, HEC = helielectric conical mesophase, K = crystal.

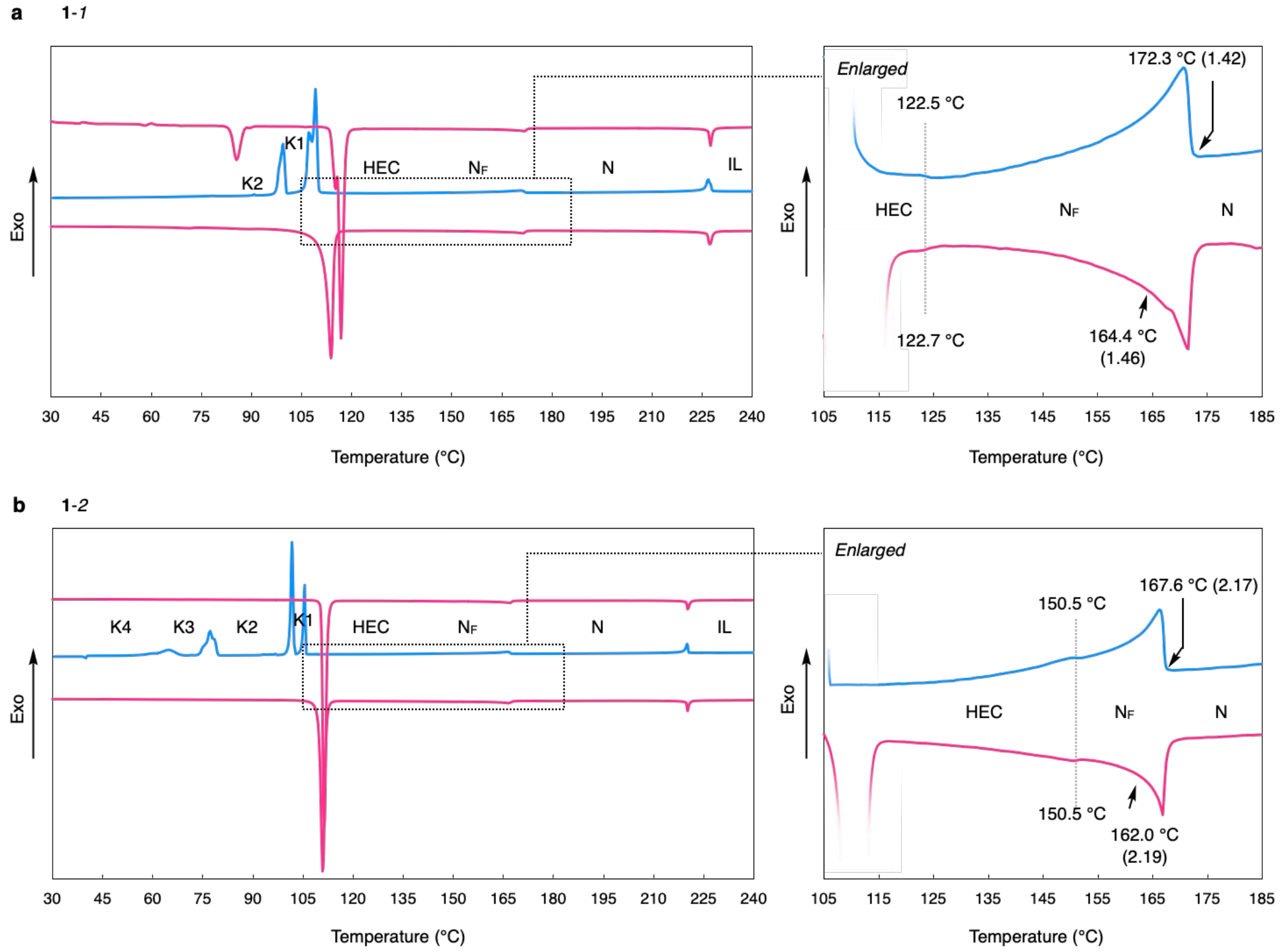

**Figure S3 |** DSC curves for **1**-*1* (a) and **1**-*2* (b). Scan rate: 10 K min$^{-1}$. Abbrev.: IL = isotropic liquid, N = nematic, $N_F$ = ferroelectric nematic, HEC = helielectric conical mesophase, K = crystal.

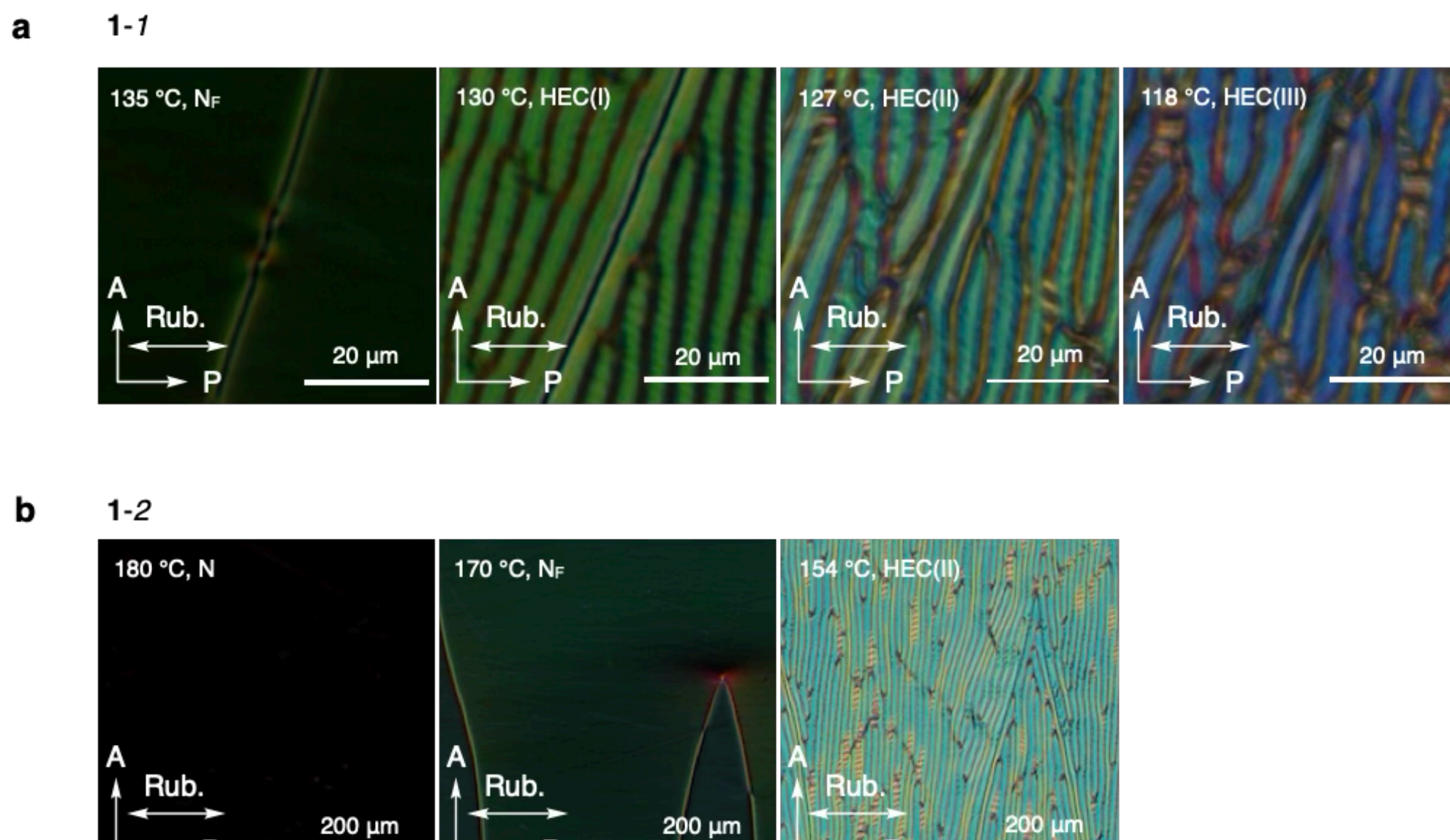

**Figure S4 |** Extra POM images for **1**-*1* (**a**) and **1**-*2* (**b**) in the antiparallel rubbed cell (thickness: 5 μm).

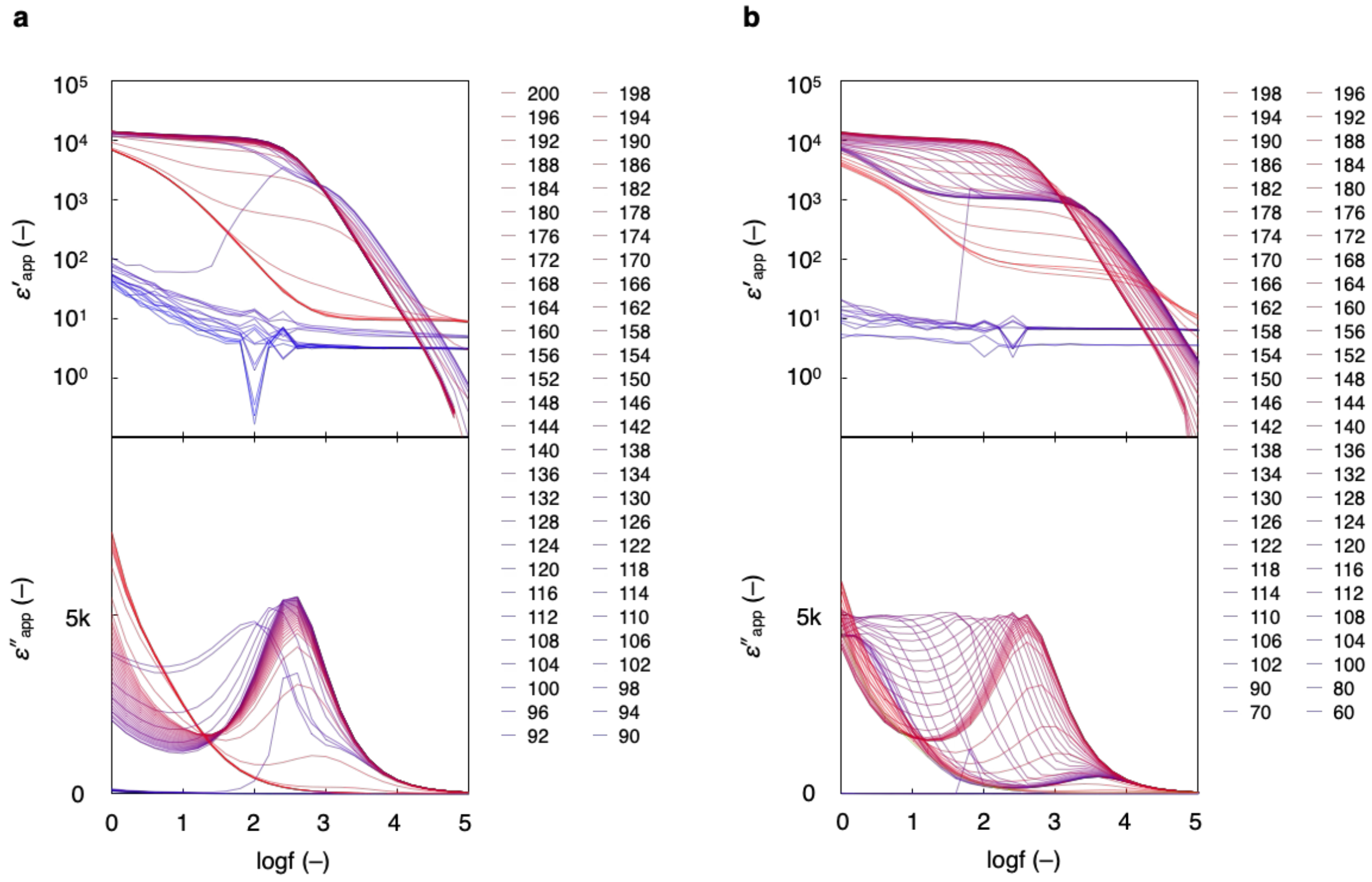

**Figure S5 |** BDS data for **1**-*1* (a) and **1**-*2* (c) in ITO glass sandwich cells (thickness: 9 μm).

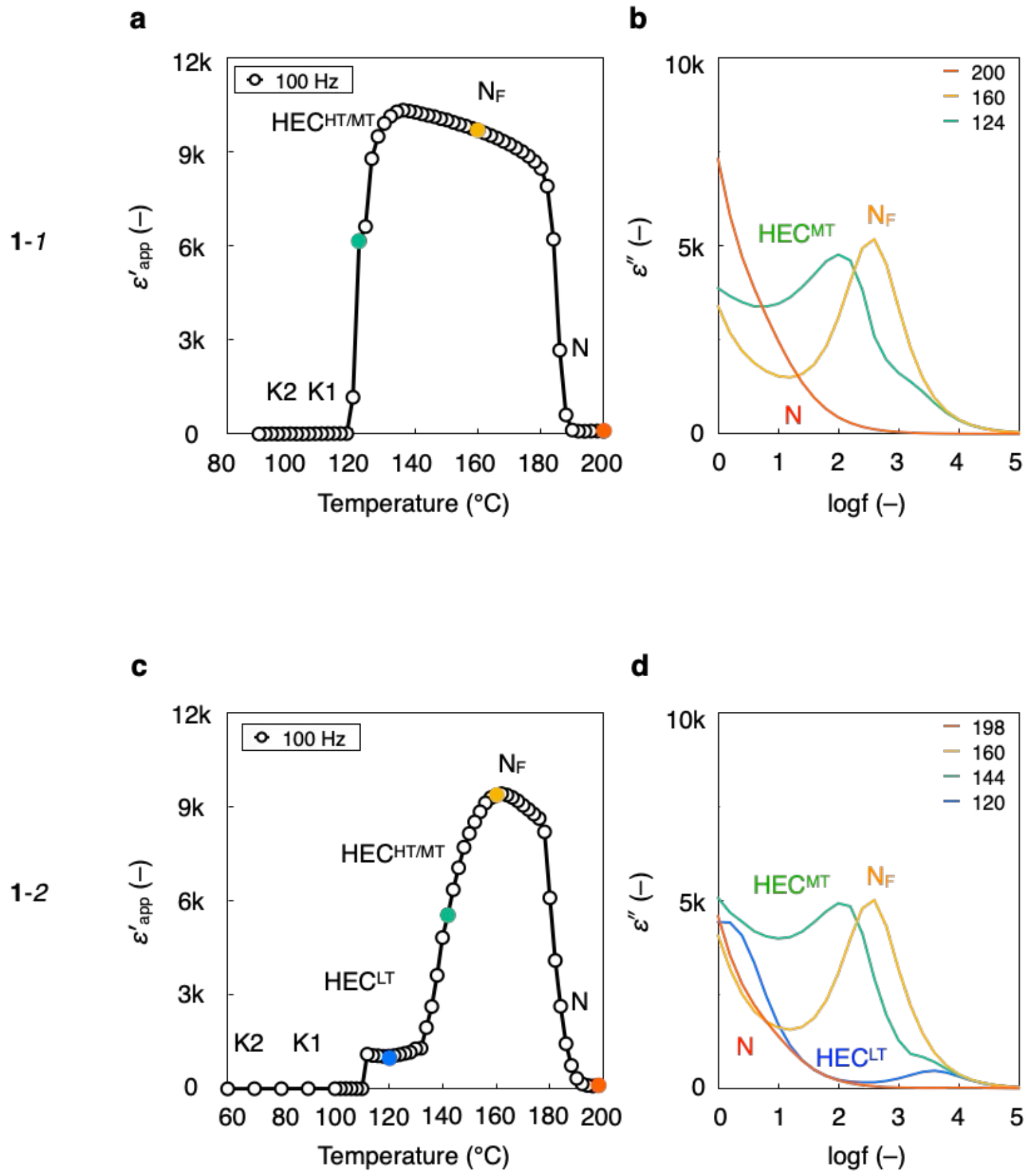

**Figure S6 |** BD data for **1-*1*** (a,b) and **1-*2*** (c,d) in ITO glass sandwich cells (thickness: 9 μm). a,c) dielectric permittivity vs temperature. b,d) dielectric loss vs frequency. The colored circles in the panel (a,c) correspond to the dielectric loss spectra in the panel (b,d), respectively.

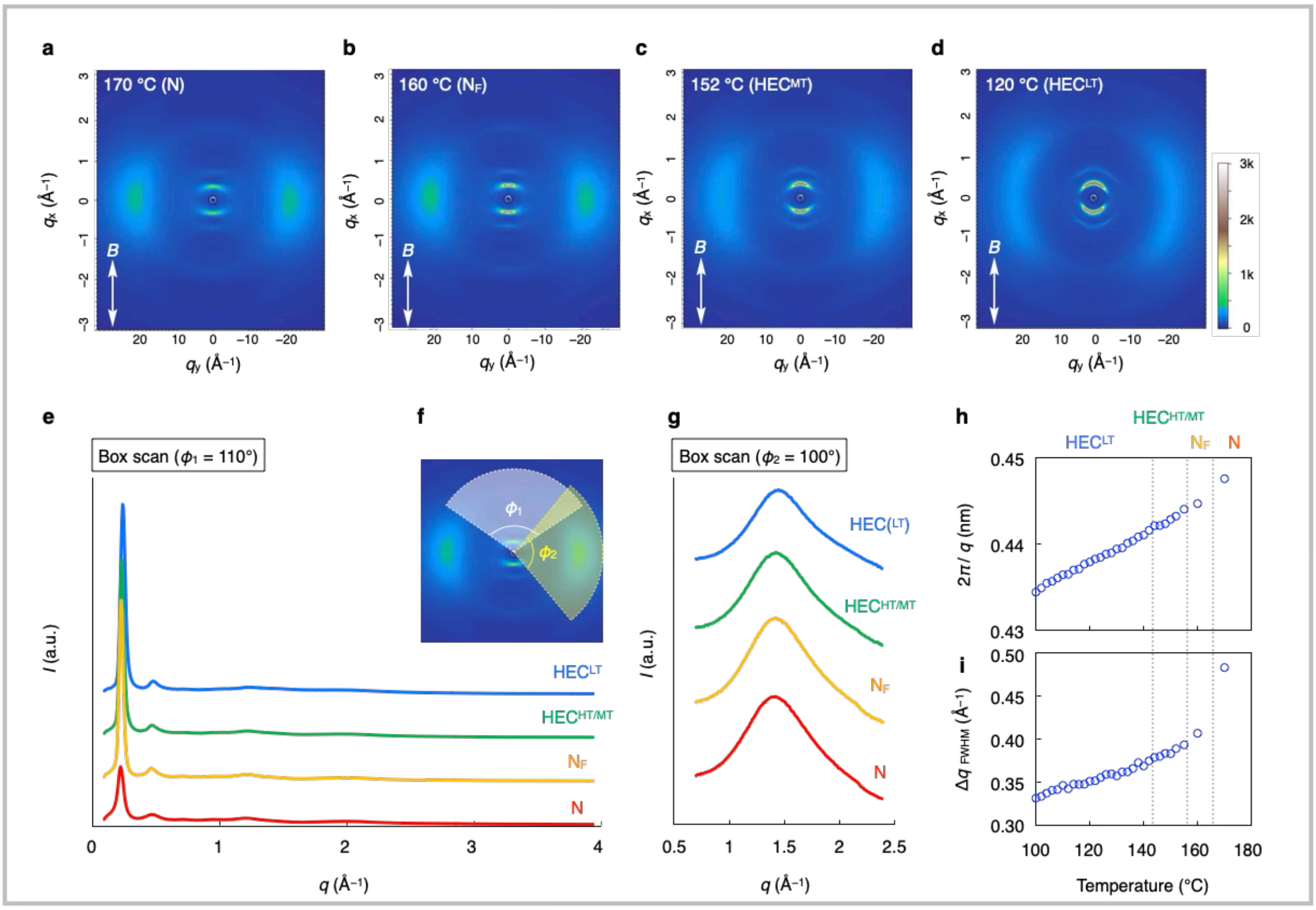

**Figure S7 |** 2D WAXD patterns under *M*-field (~1 T) in the N (a), $N_F$ (b), $HEC^{MT}$ (c) and $HEC^{LT}$ (d) phases. Symbol *B* denote the direction of *M*-field. 1D X-ray diffractograms along meridional direction (**n** ∥ **B**) (e) and equatorial direction (**n** ⊥ **B**) (g). f) Box scan's area ($\varphi_1$ = 110°, $\varphi_2$ = 100°). h) $2\pi/q$ vs Temperature. **i** $\Delta q_{FWHM}$ vs Temperature.

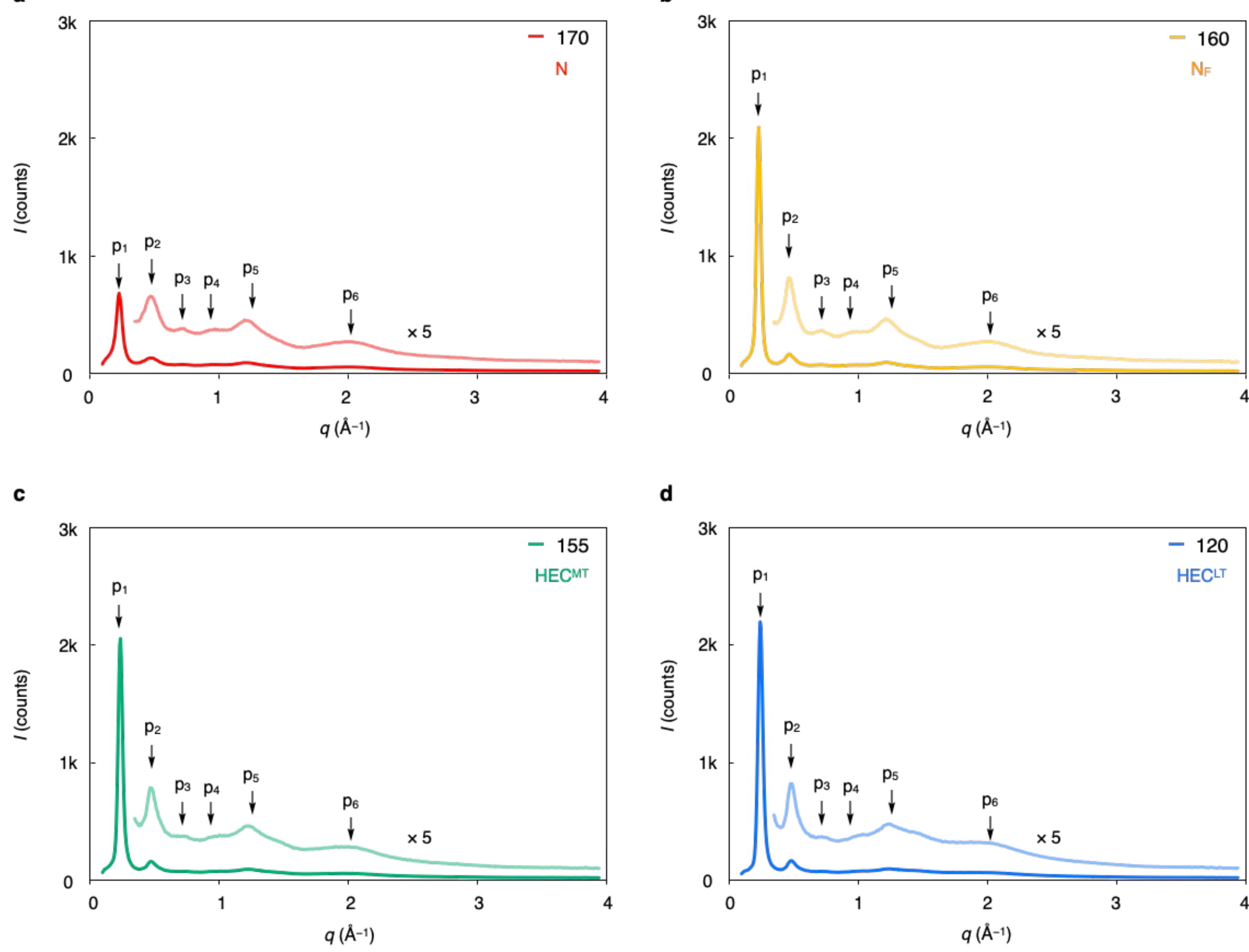

**Figure S8 |** 1D WAXD patterns under *M*-field (~1 T) in the N (a), $N_F$ (b), $HEC^{MT}$ (c) and $HEC^{LT}$ (d) phases.

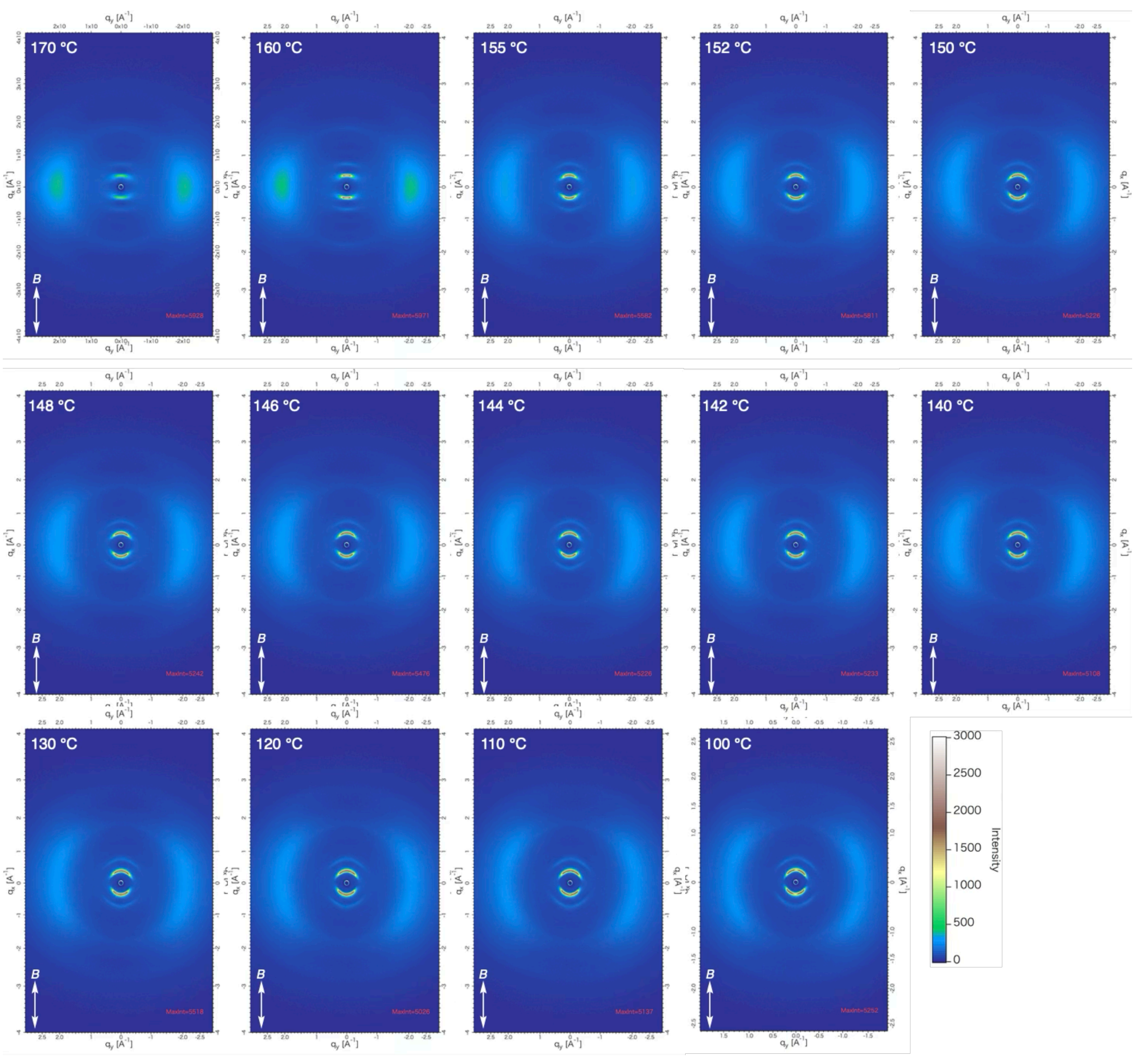

**Figure S9 |** 2D WAXD patterns under *M*-field (~1 T) for **1-*3***.

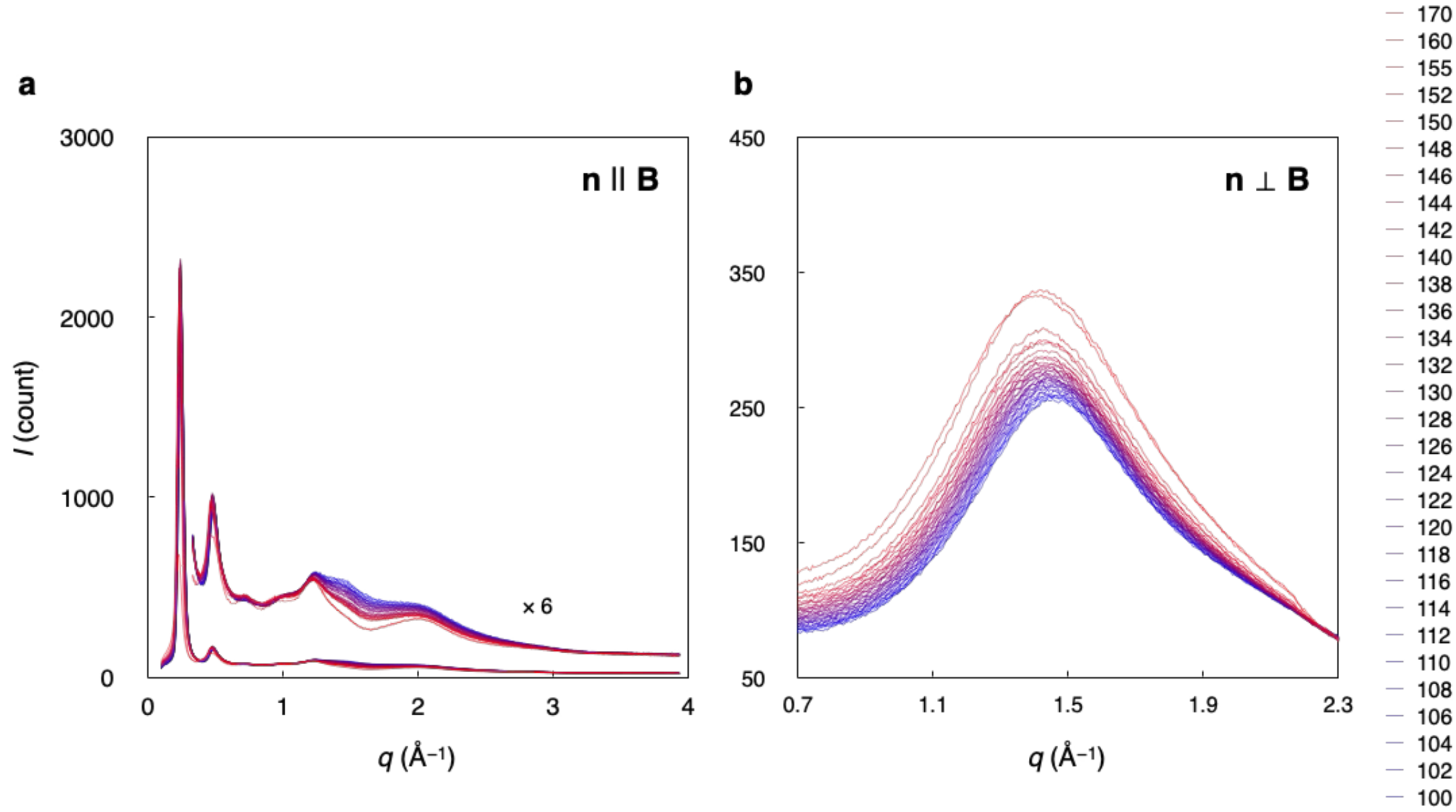

**Figure S10 |** Temperature dependence of 1D wide-angle X-ray diffractogram analyzed along meridional (**a**) and equatorial (**b**) direction in the corresponding 2D WAXD patterns (Figure S9) for **1-*3***.

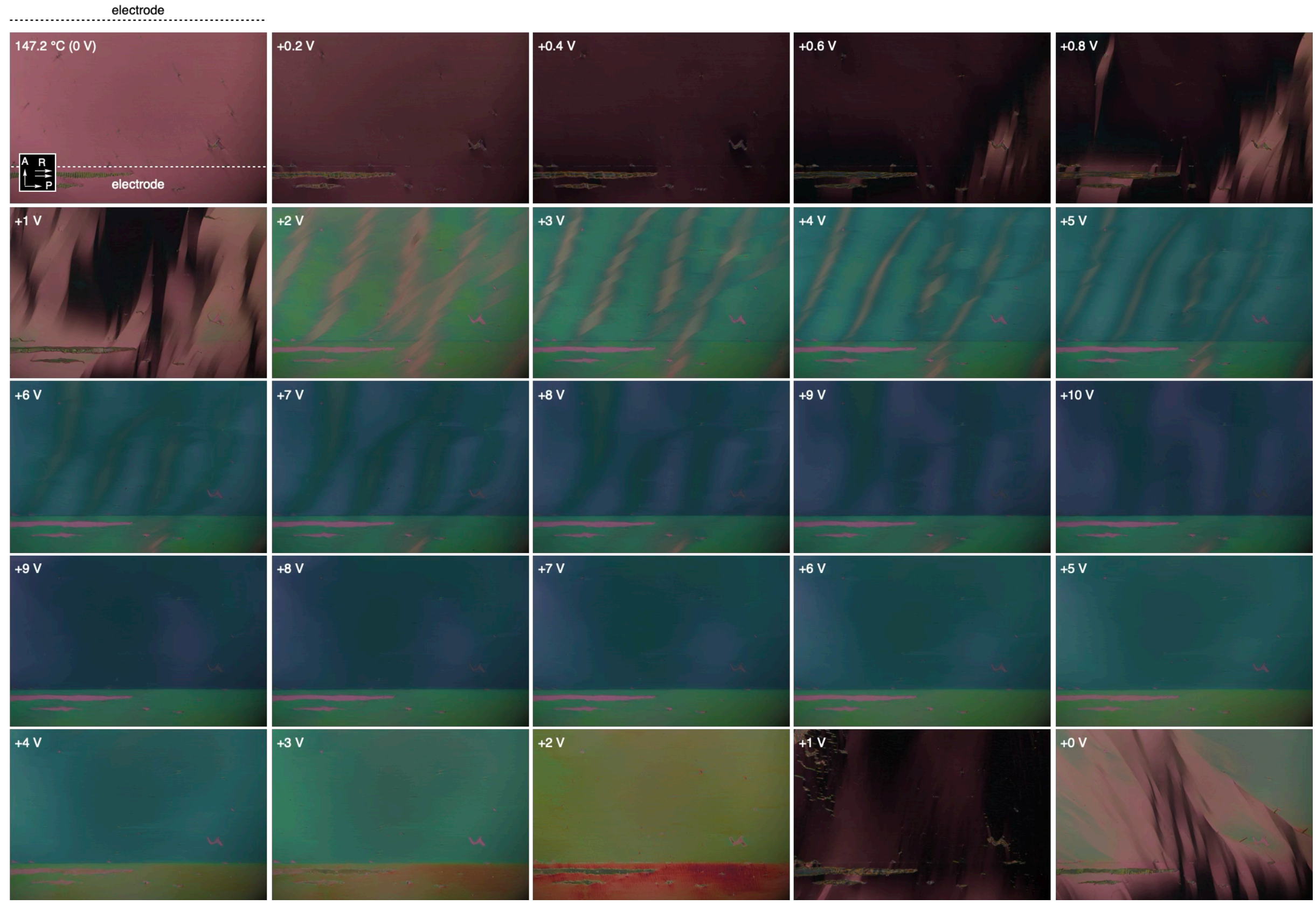

**Figure S11 |** POM texture changes for **1**-*3* in the synparallel cell (thickness: 10 µm) under *E*-field from 0 to +1 to 0 (again) V mm$^{-1}$ (**R** ⊥ **E**).

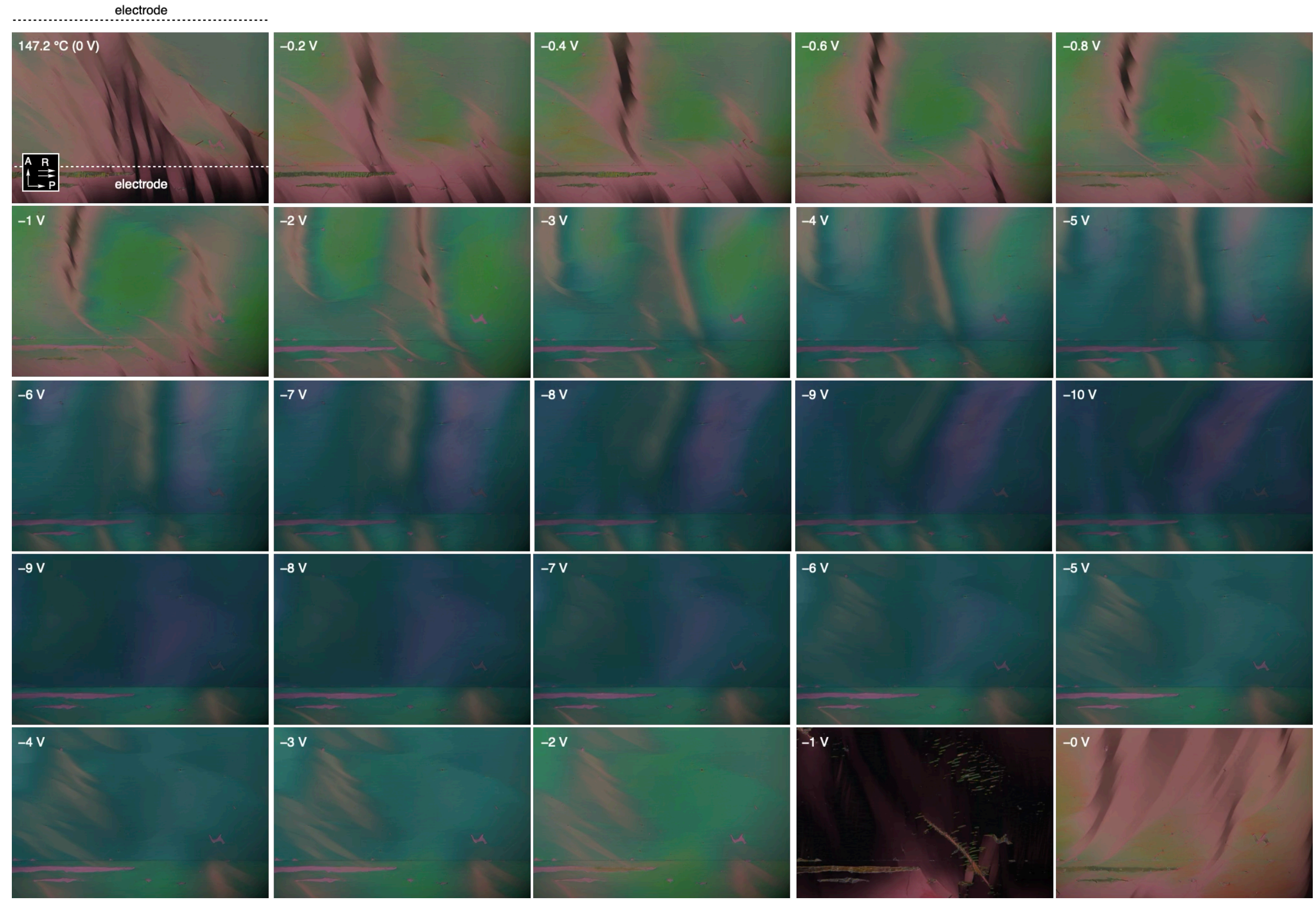

**Figure S12 |** POM texture changes for **1**-*3* in the synparallel cell (thickness: 10 μm) under *E*-field from 0 to −1 to 0 (again) V $mm^{-1}$ (**R** ⊥ **E**).

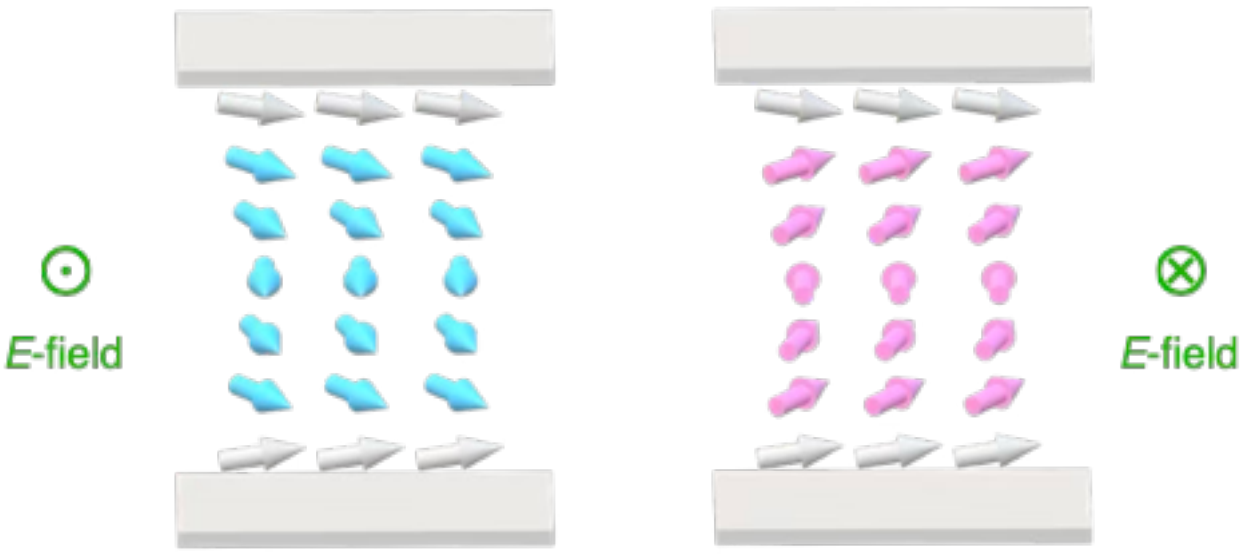

**Figure S13 |** Schematic illustration of director field in the SP cell (**R** ⊥ **E**) under *E*-field.

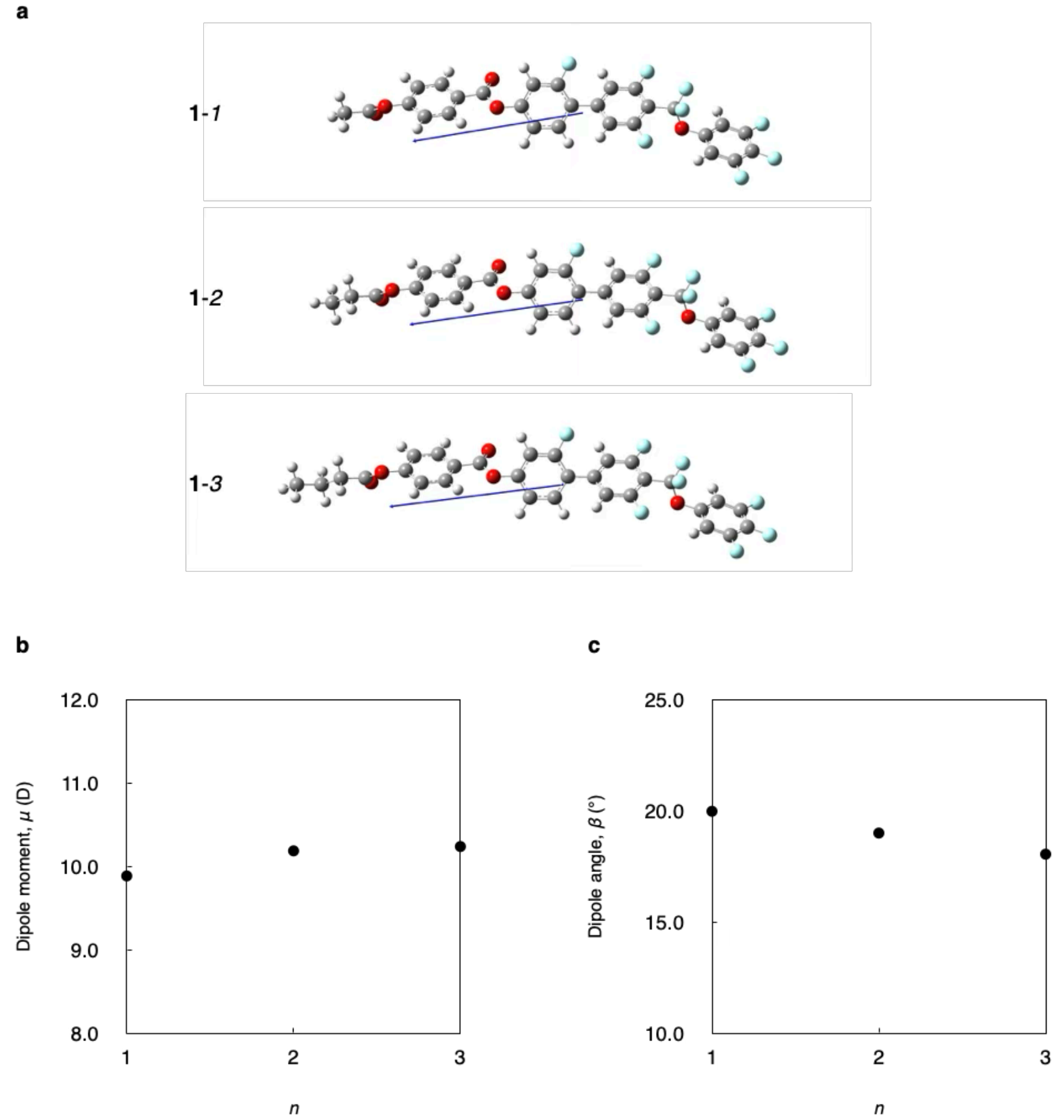

**Figure S14 |** Results of DFT calculation for **1**-*n* (n = 1–3): a) optimized structures; b) dipole moment ($\mu$) vs *n*; c) dipole angle ($\beta$) vs *n*.

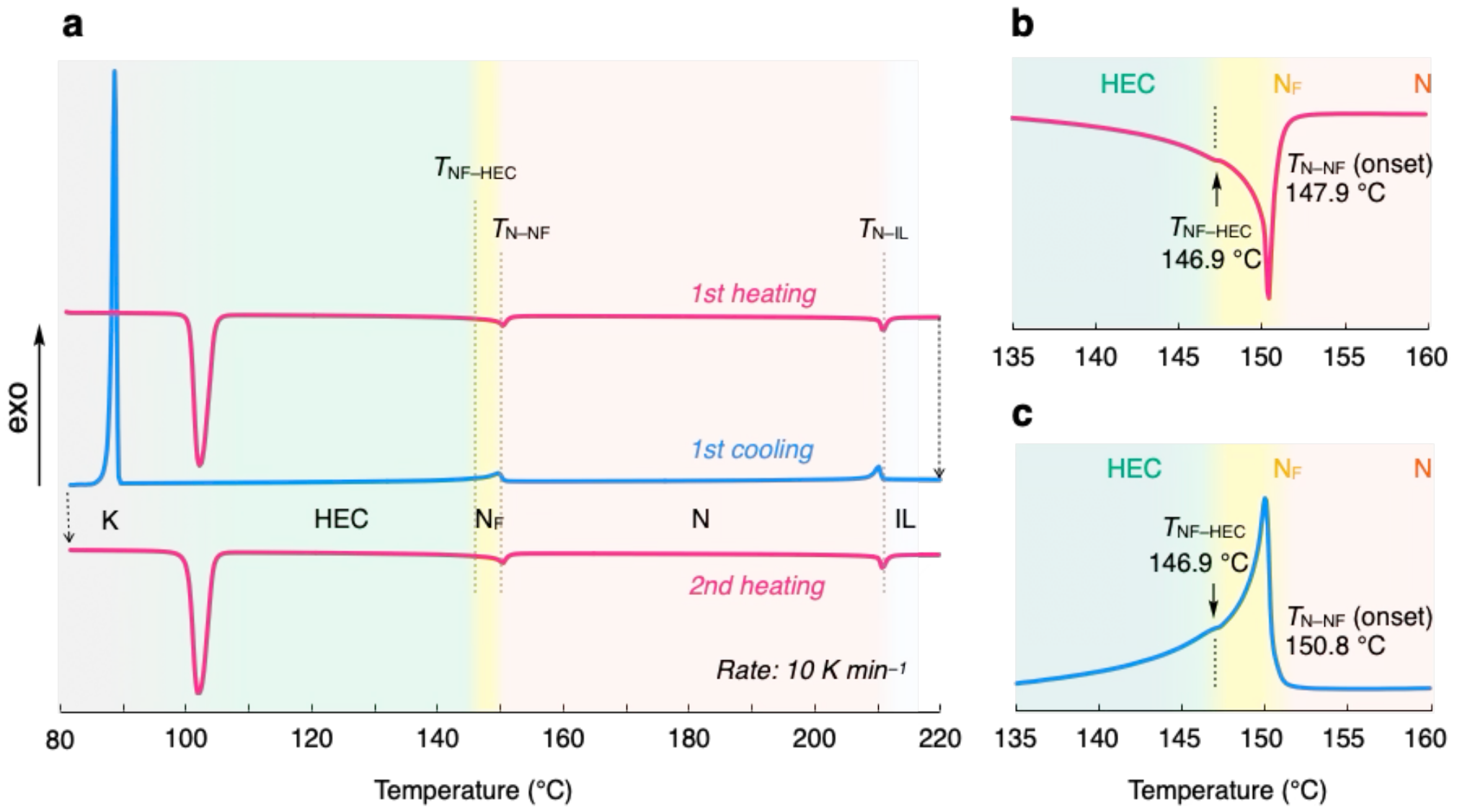

**Figure S15 |** Completed DSC data for **1**-*3*.

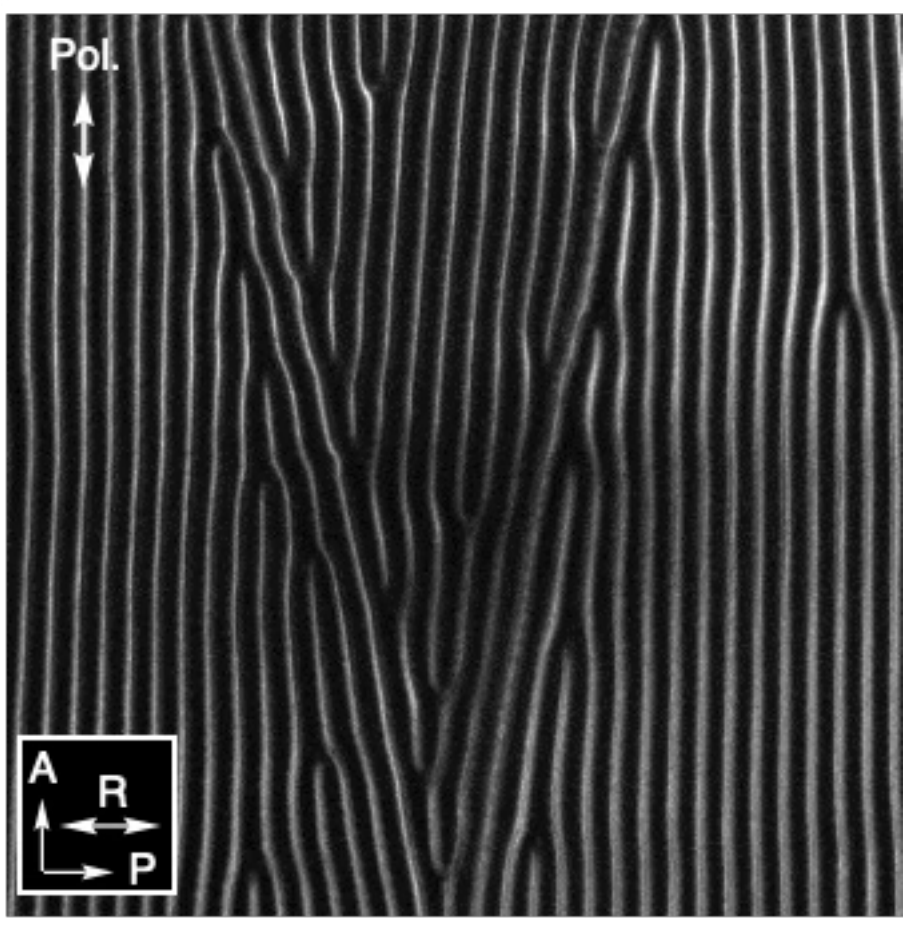

**Figure S16 |** SH micrography for **1**-*3* in an AP cell (5 μm). Pol. denotes a polarized plane.

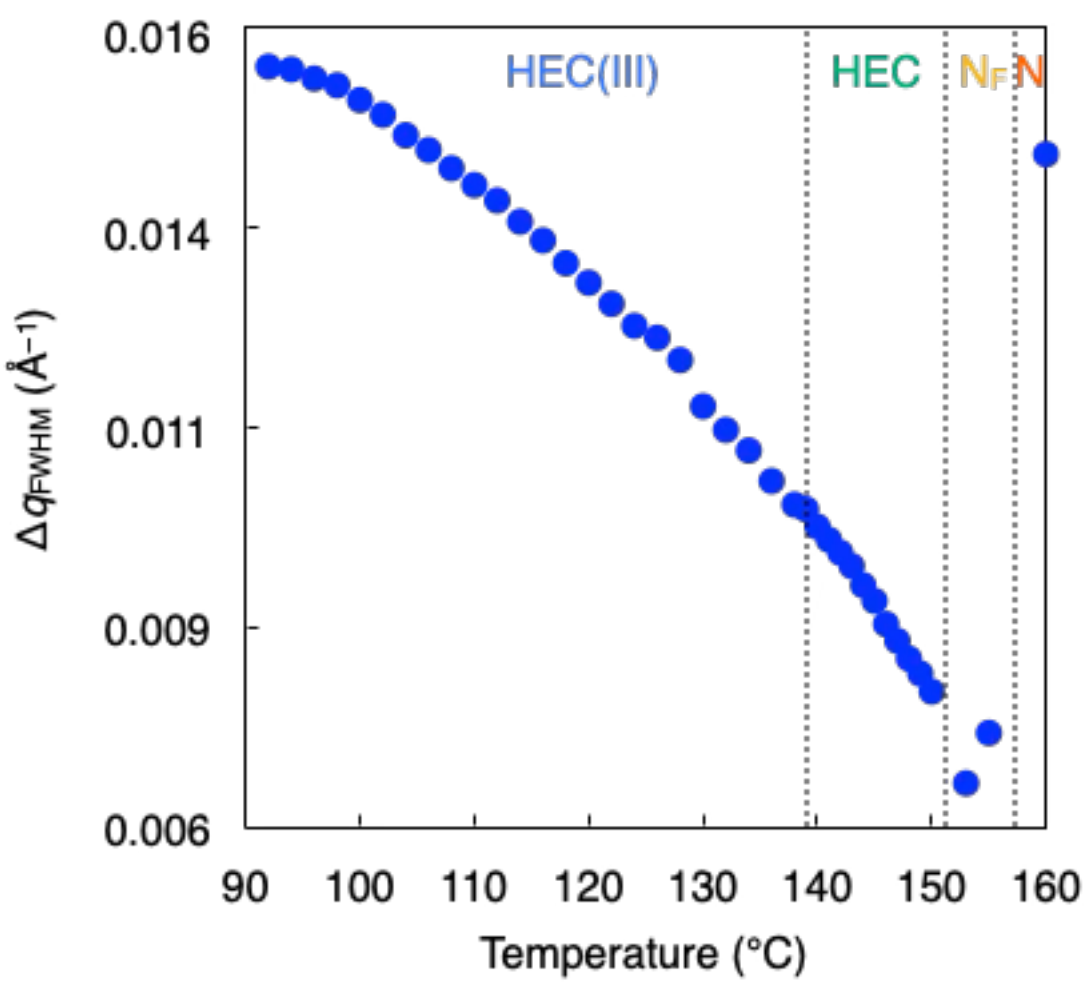

**Figure S17 |** $\Delta q_{\mathrm{FWHM}}$ vs Temperature for **1**-*3*.

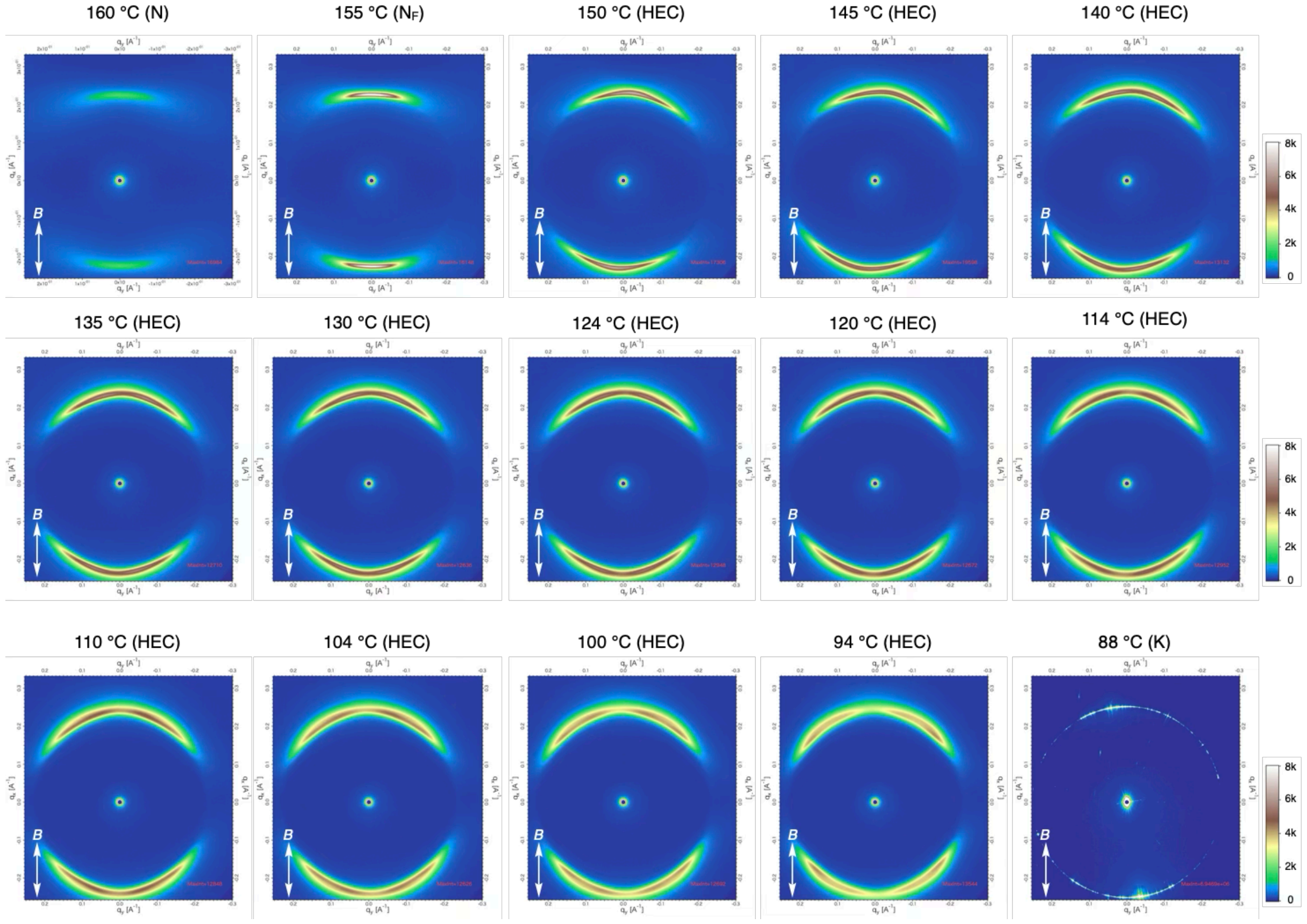

**Figure S18 |** 2D SAXS patterns under *M*-field (~1 T) for **1-*3***.

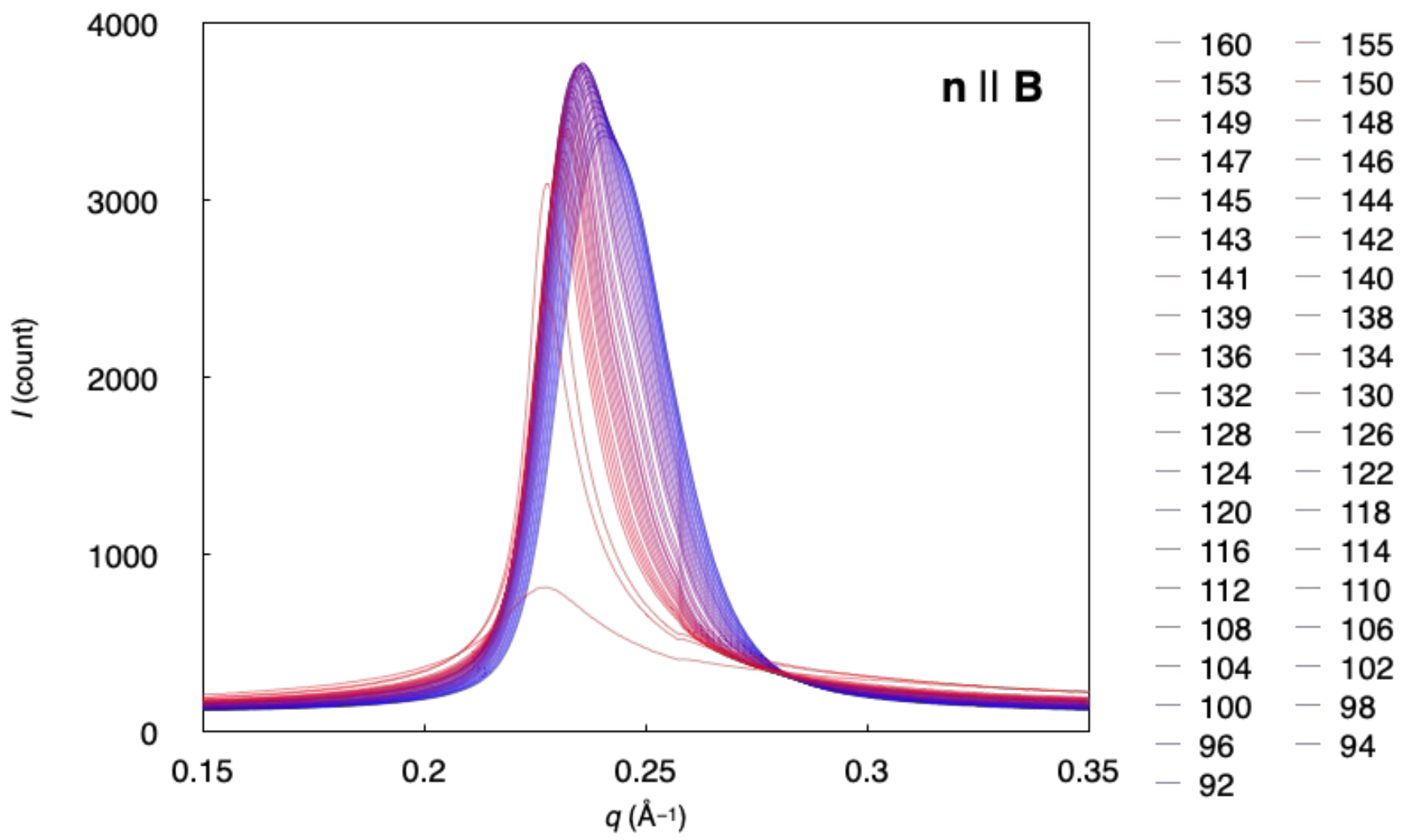

**Figure S19 |** Temperature dependence of 1D small-angle X-ray diffractogram for **1-*3***.

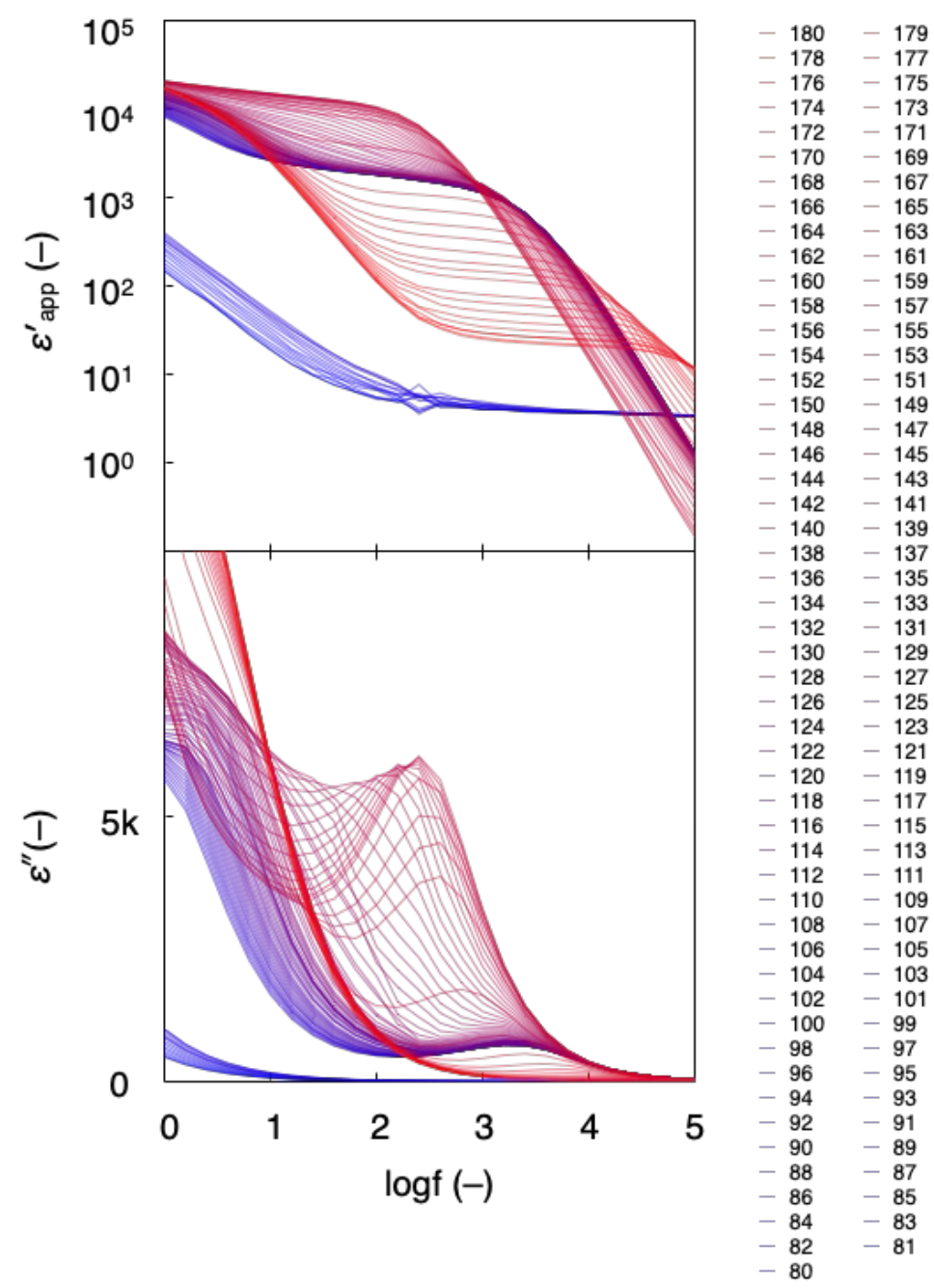

**Figure S20 |** 2D BDS spectra for **1**-*3* in various temperatures.

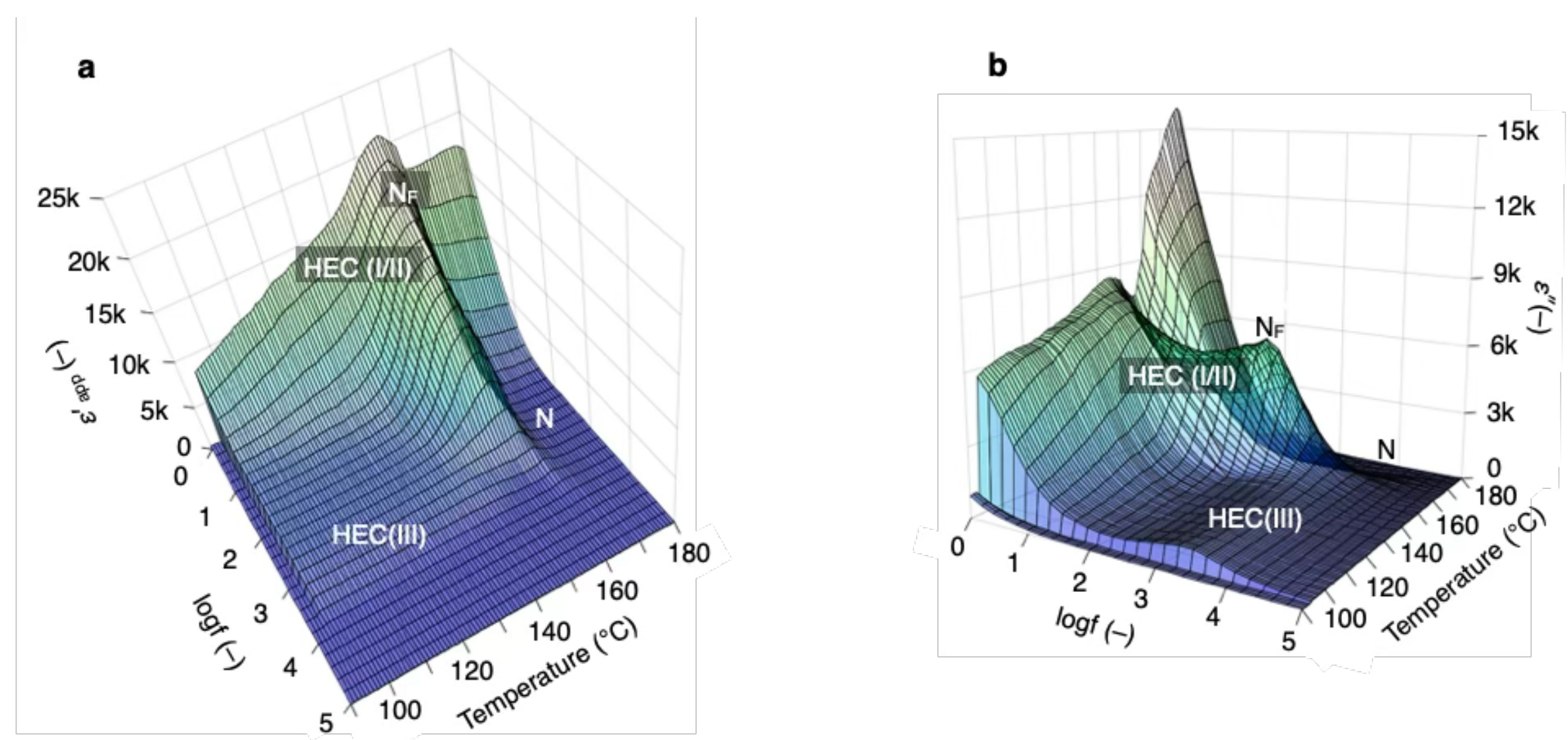

**Figure S21 |** 3D BDS spectra for **1**-*3* in various temperatures. a) $\varepsilon'_{app}$ vs log*f*, Temperature; b) $\varepsilon''$ vs log*f*, Temperature.

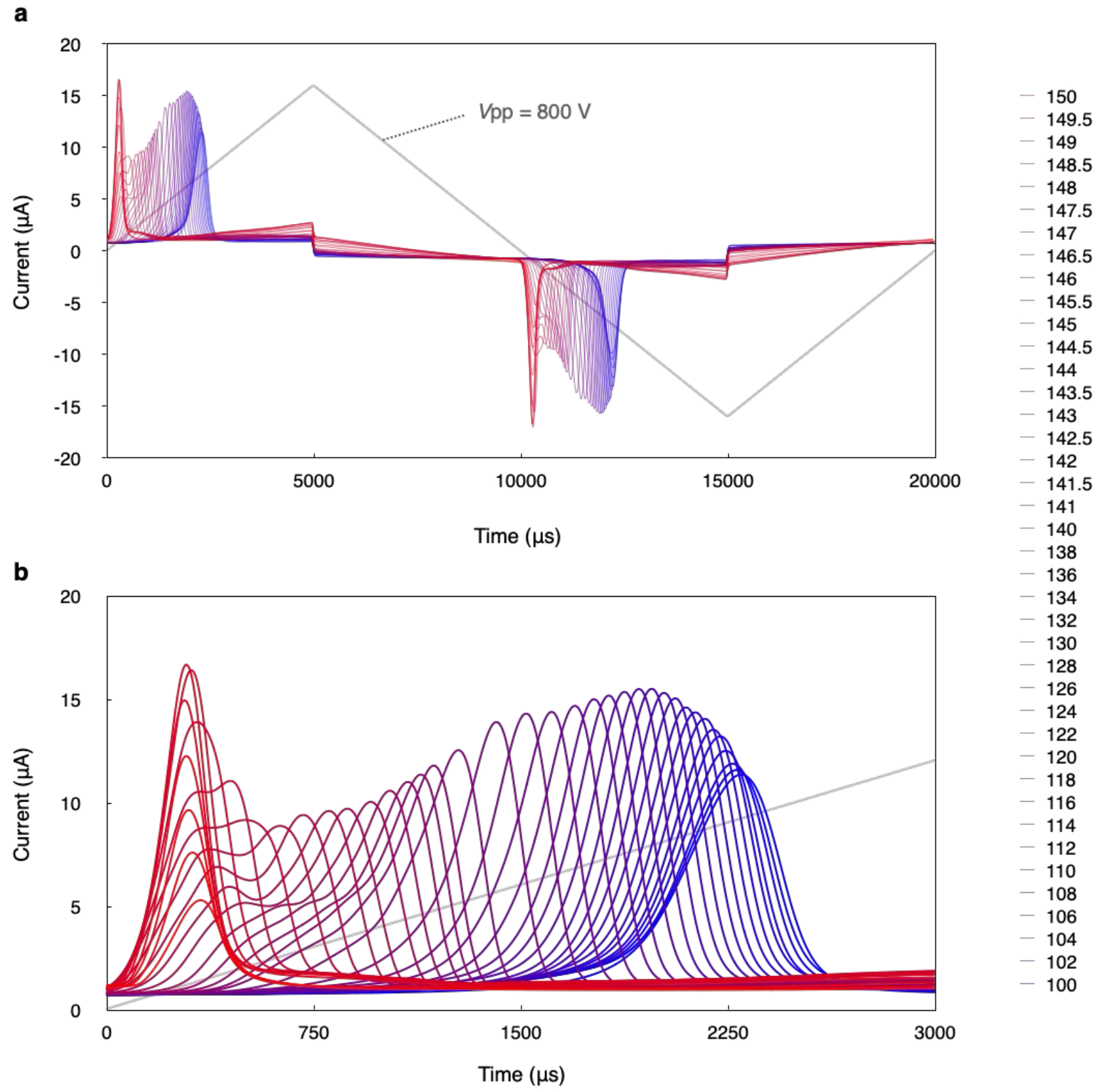

**Figure S22 |** PRC profiles for **1**-*3* under *E*-field (triangular wave, 50 Hz, 800 $V_{pp}$) in the nonparallel rubbed cell (thickness: 5.6 μm, electrode distance: 0.5 mm).

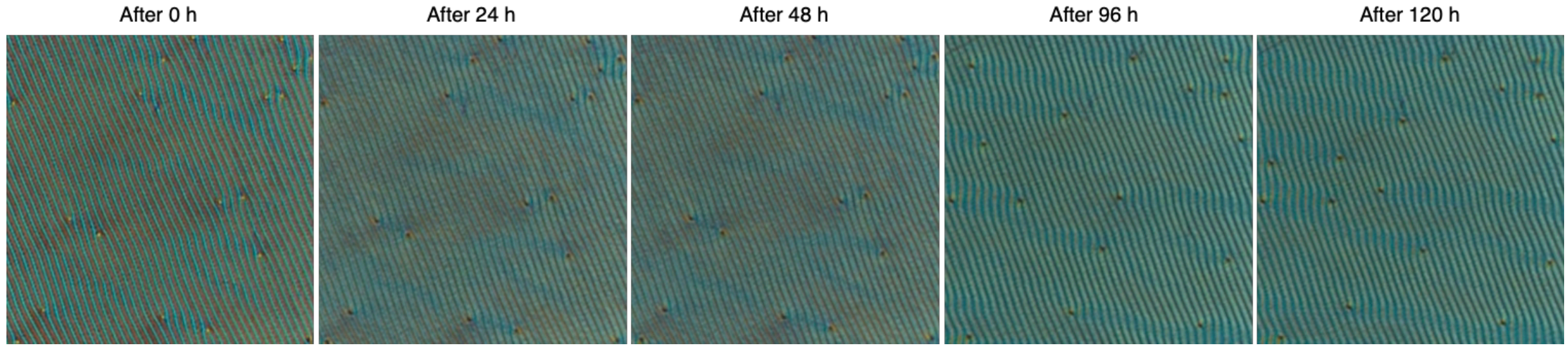

**Figure S23 |** Time dependence of POM texture changes in the tilted stripes after removal *E*-field.

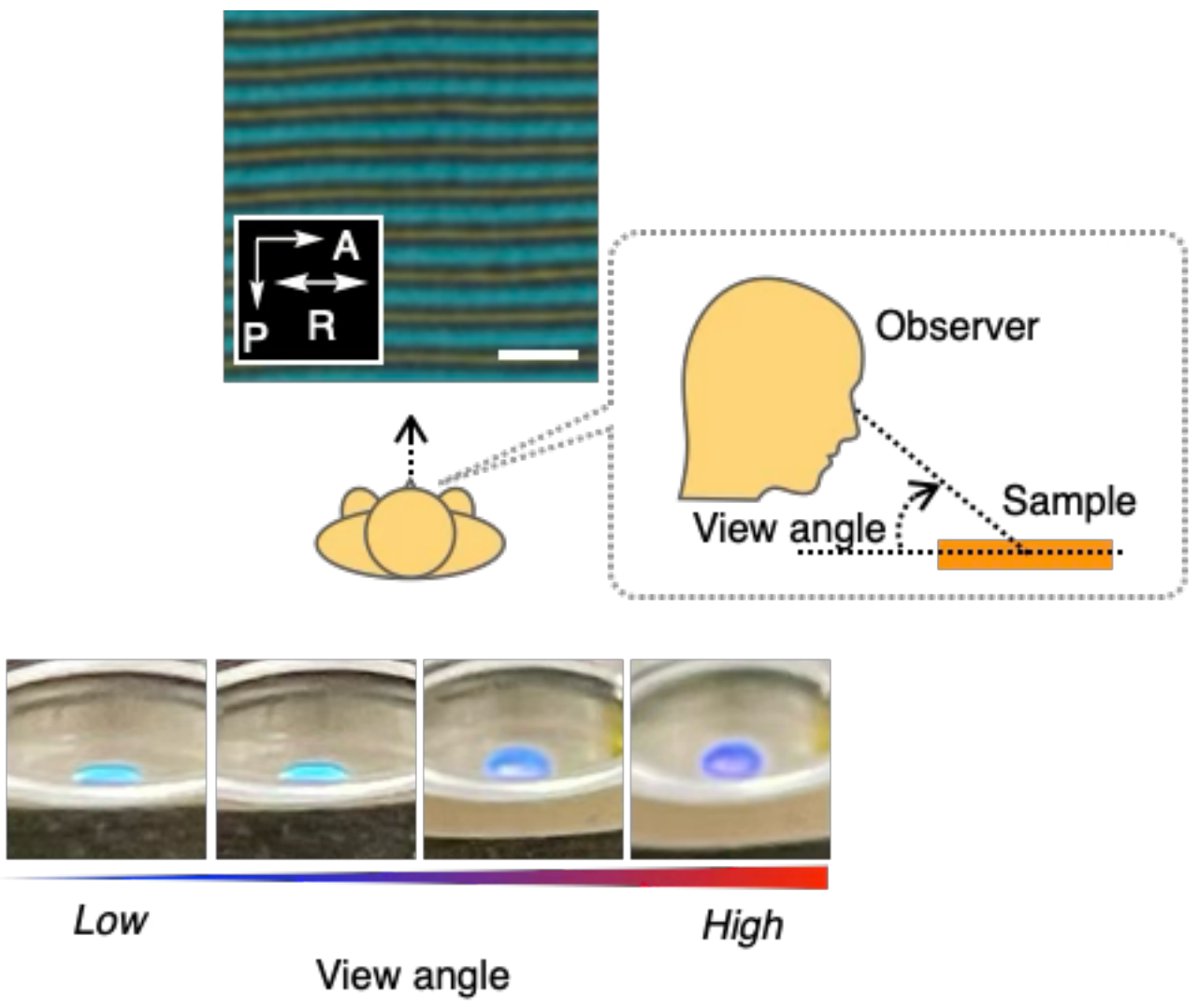

**Figure S24 |** View angle dependence of the color of diffraction pattern. At large view angles, the color blue-shifted from sky-blue to purple.

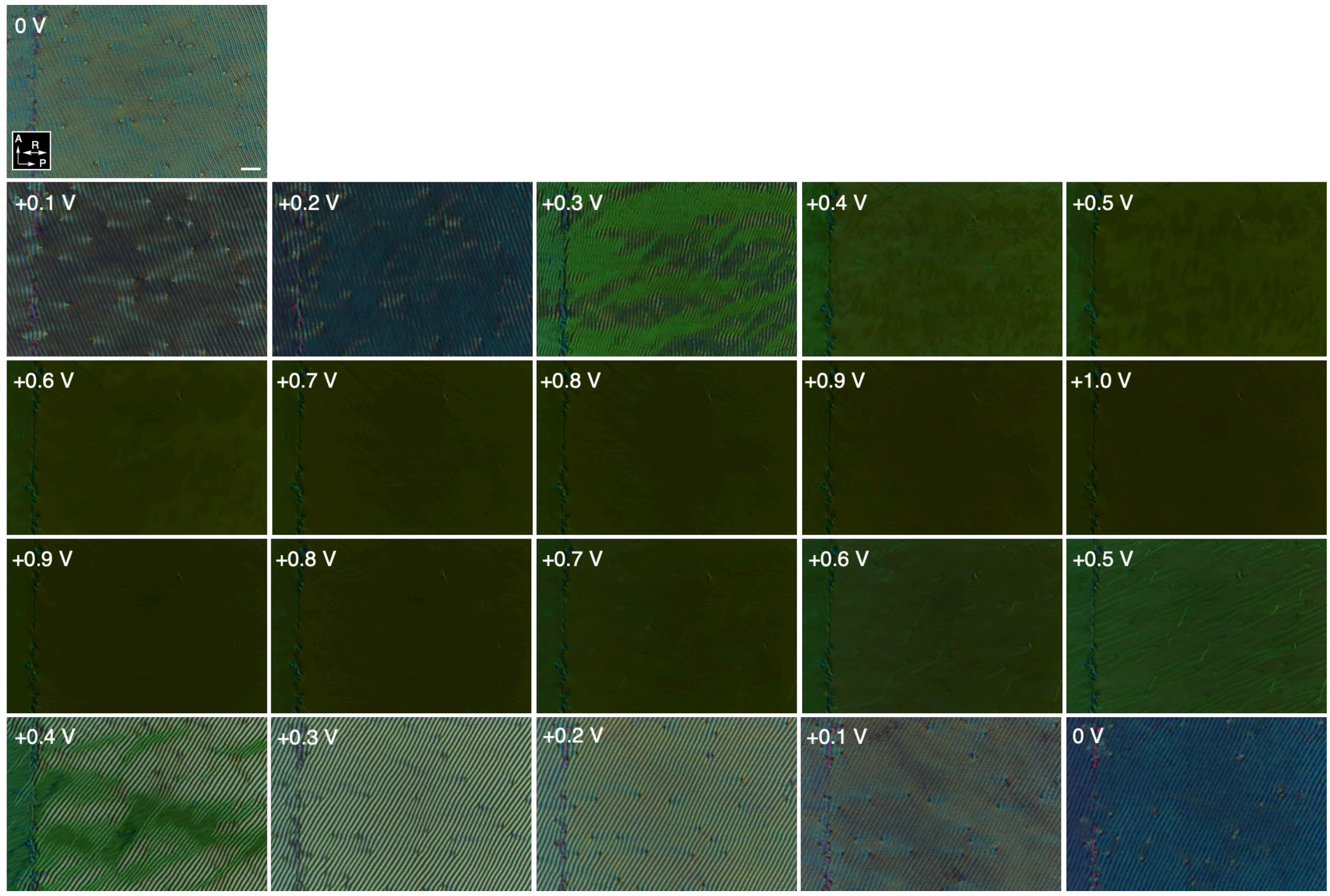

**Figure S25 |** POM texture changes for **1**-*3* in the antiparallel cell (thickness: 5 μm) under *E*-field from 0 to +1 to 0 (again) V $mm^{-1}$ (**R** ∥ **E**). Scale bar: 20 μm.

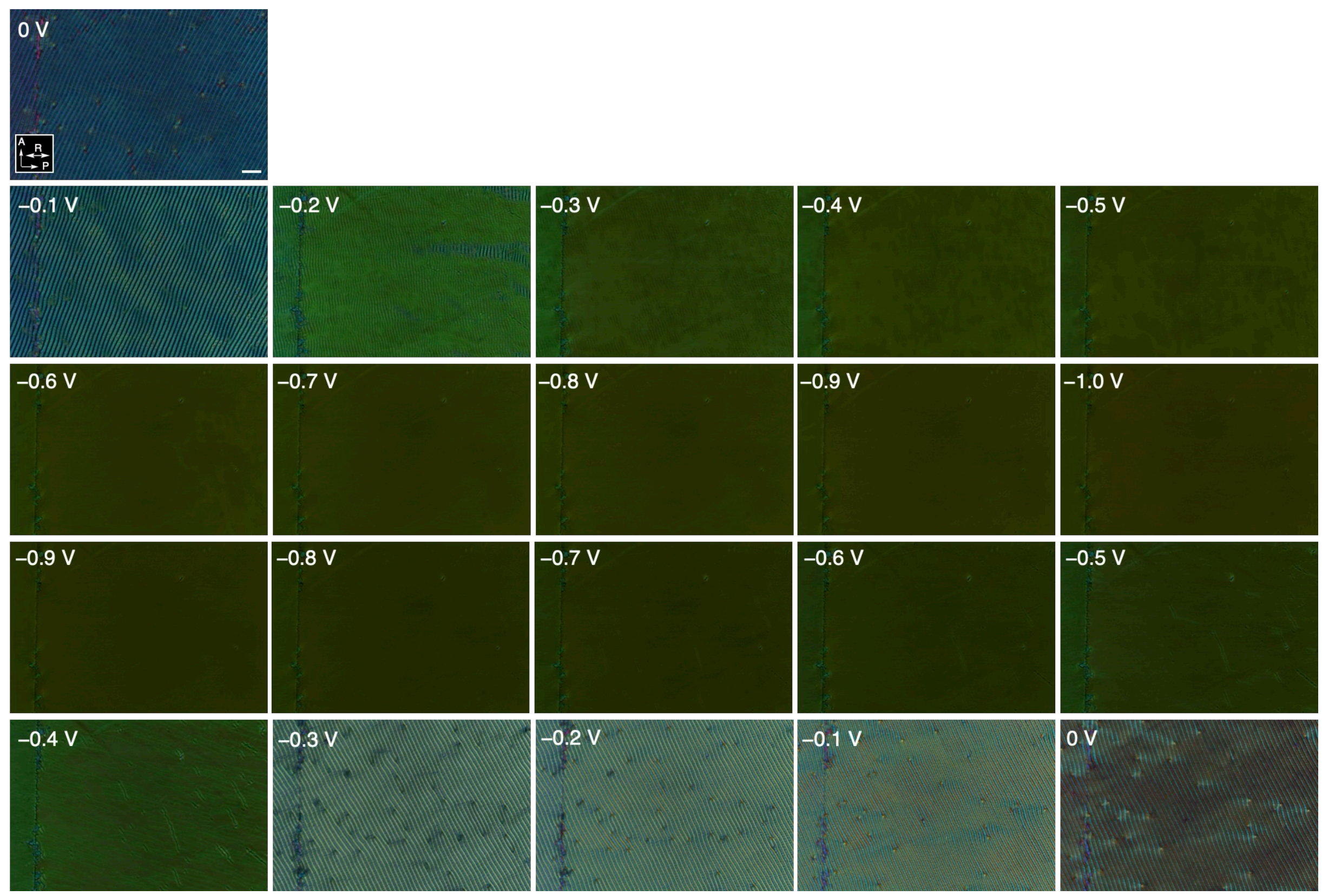

**Figure S26 |** POM texture changes for **1**-*3* in the antiparallel cell (thickness: 5 μm) under *E*-field from 0 to −1 to 0 (again) V mm$^{-1}$ (**R** ∥ **E**). Scale bar: 20 μm.

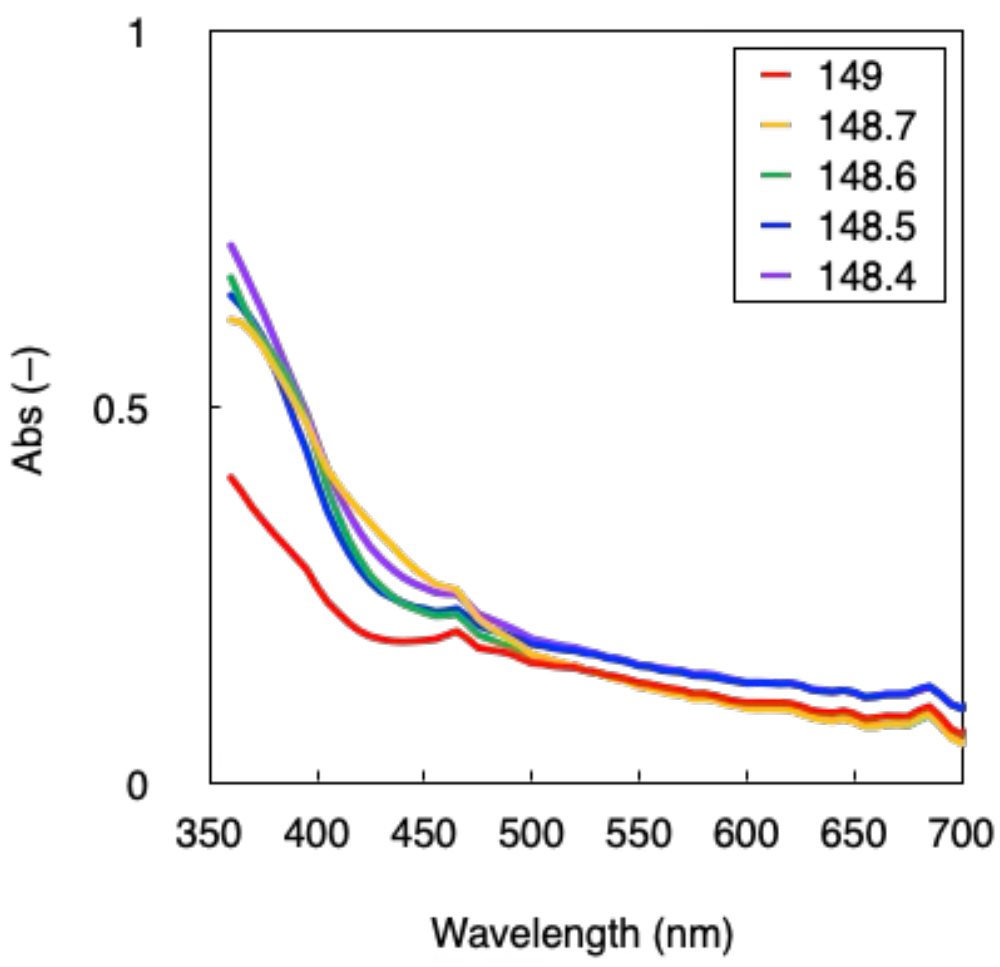

**Figure S27 |** Absorbance spectra recorded by CD spectroscopy for **1**-*3* in the antiparallel cell (thickness: 10 μm) under *E*-field of +1 V mm$^{-1}$ (**R** ⊥ **E**). The magnitude of absorbance is sufficiently small. Thus, the obtained CD spectra is not artifact. For all spectra, their shape and position are almost same. Therefore, CD peak shift in the panel (f) in Figure 6 (original manuscript) is not artifact.

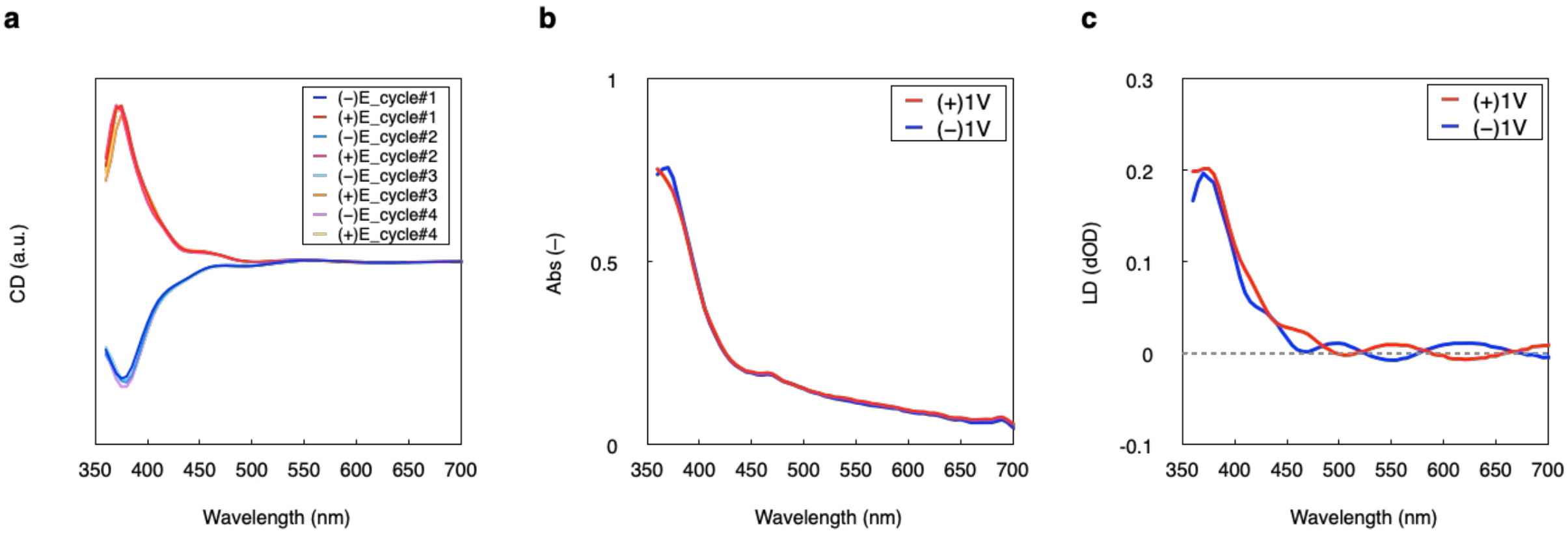

**Figure S28 |** Data recorded by CD spectroscopy for **1**-*3* in the antiparallel cell (thickness: 10 μm) under *E*-field of ±1 V mm$^{-1}$ (**R** ⊥ **E**). a) CD intensity vs wavelength , b) absorbance vs wavelength, c) linear dichroism (LD) vs wavelength. Note: The magnitude of absorbance and LD are sufficiently small. Thus, the obtained CD spectra is not artifact.

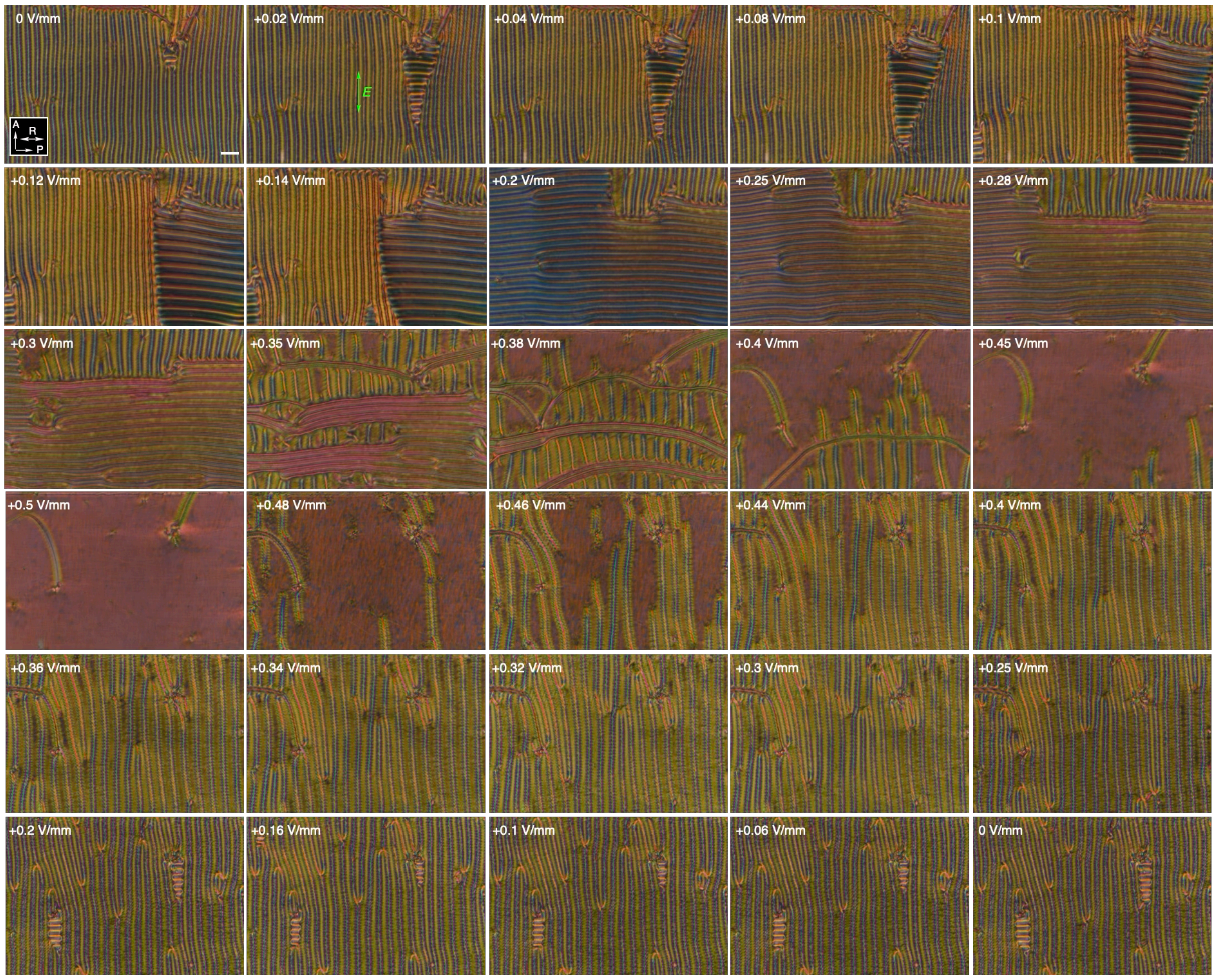

**Figure S29 |** POM texture changes for **1**-*3* in the antiparallel cell (thickness: 10 μm) under *E*-field from 0 to +0.5 to 0 (again) V mm$^{-1}$ (**R** ⊥ **E**). Scale bar: 20 μm.

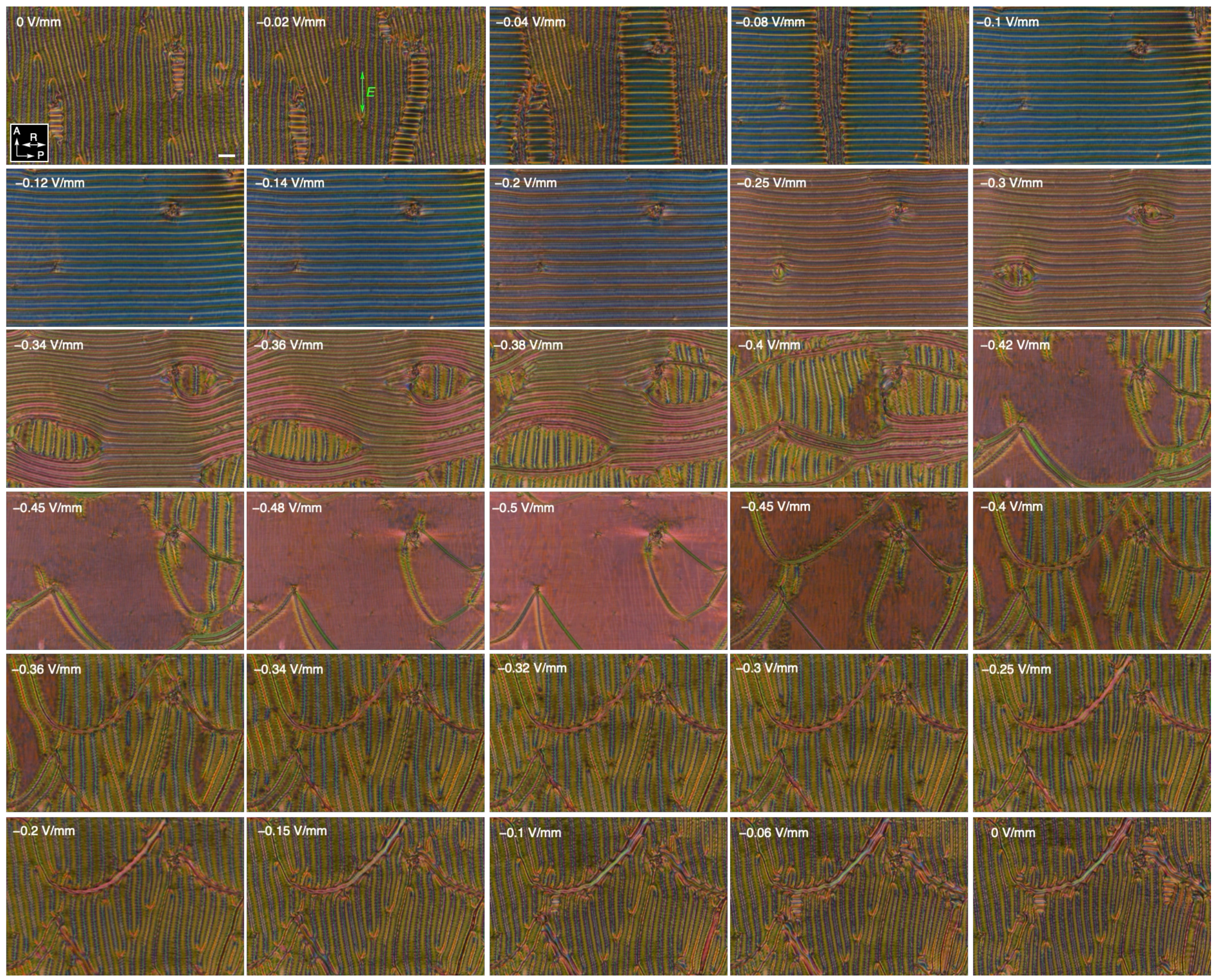

**Figure S30 |** POM texture changes for **1**-*3* in the antiparallel cell (thickness: 10 μm) under *E*-field from 0 to −0.5 to 0 (again) V mm$^{-1}$ (**R** ⊥ **E**). Scale bar: 20 μm.

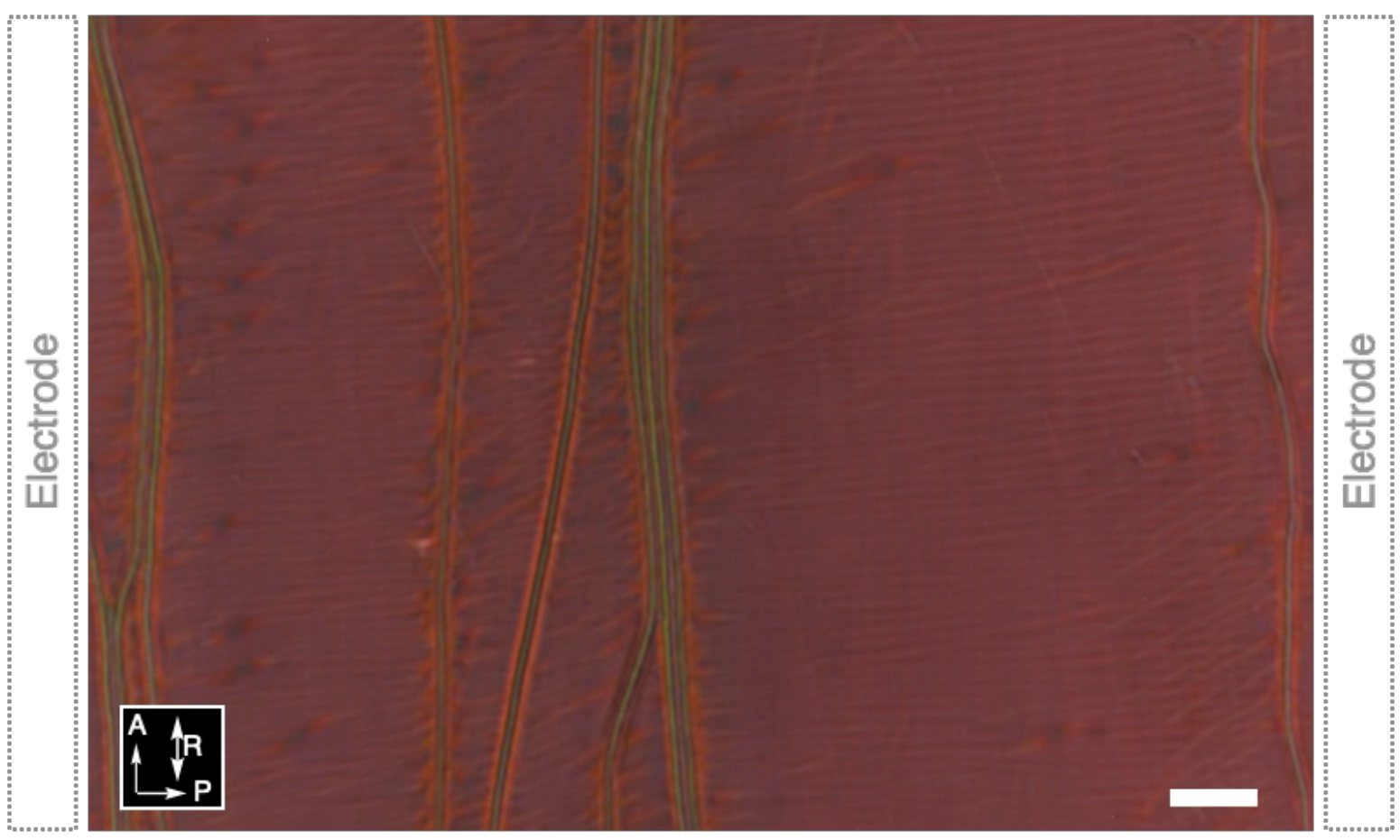

**Figure S31 |** POM texture changes for **1**-*3* in the antiparallel cell (thickness: 10 μm) under *E*-field of −0.5 V $mm^{-1}$ (**R** ⊥ **E**). Scale bar: 20 μm.

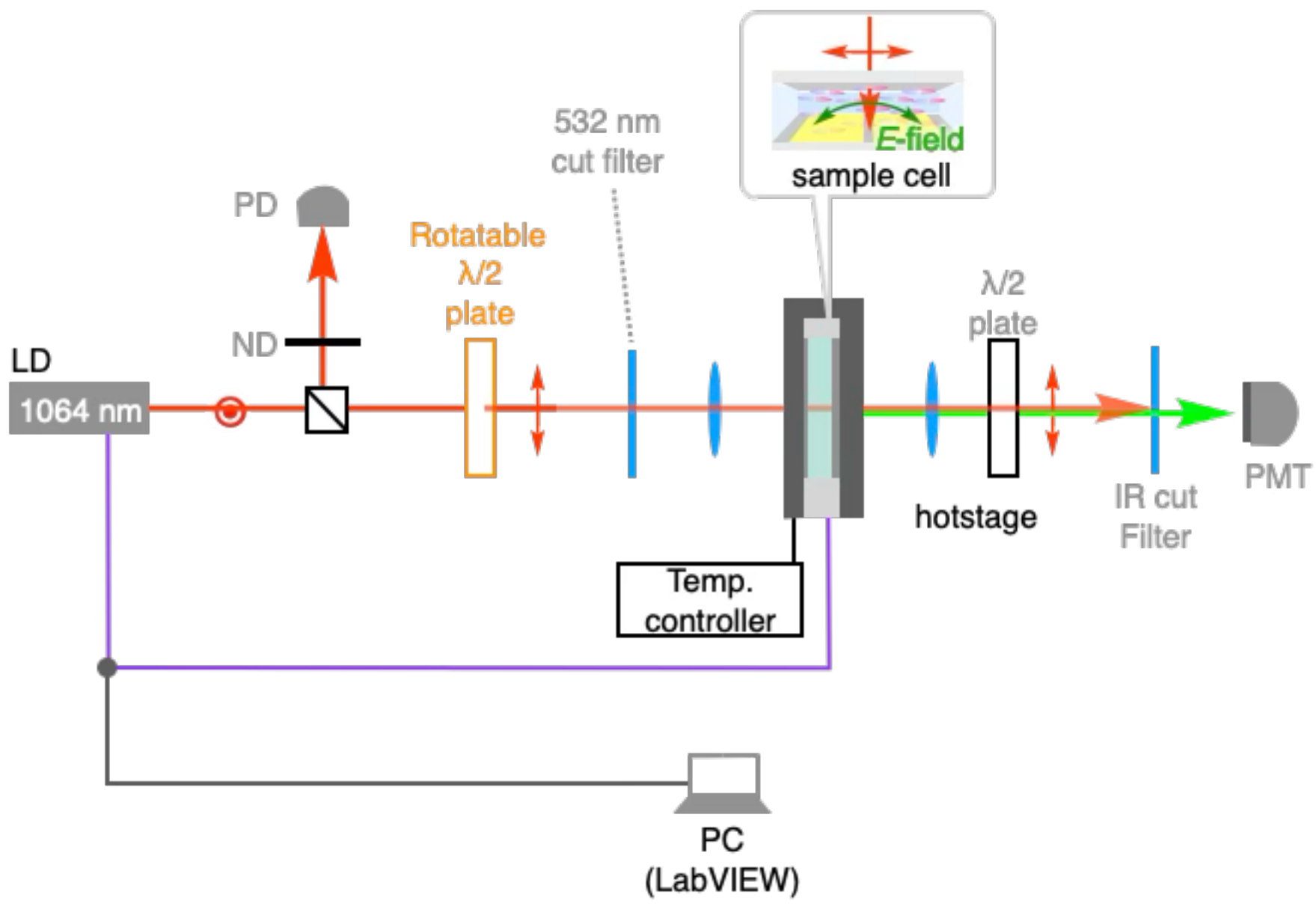

**Figure S32 |** Optical setup for SHG studies.

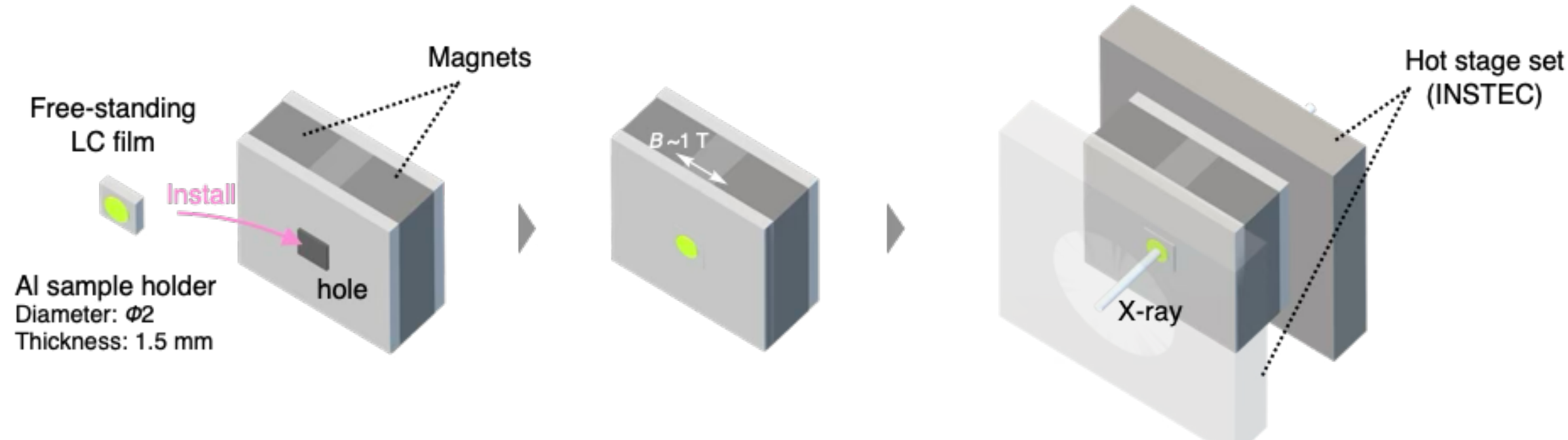

**Figure S33 |** Measurement setup for XRD studies under magnetic field.

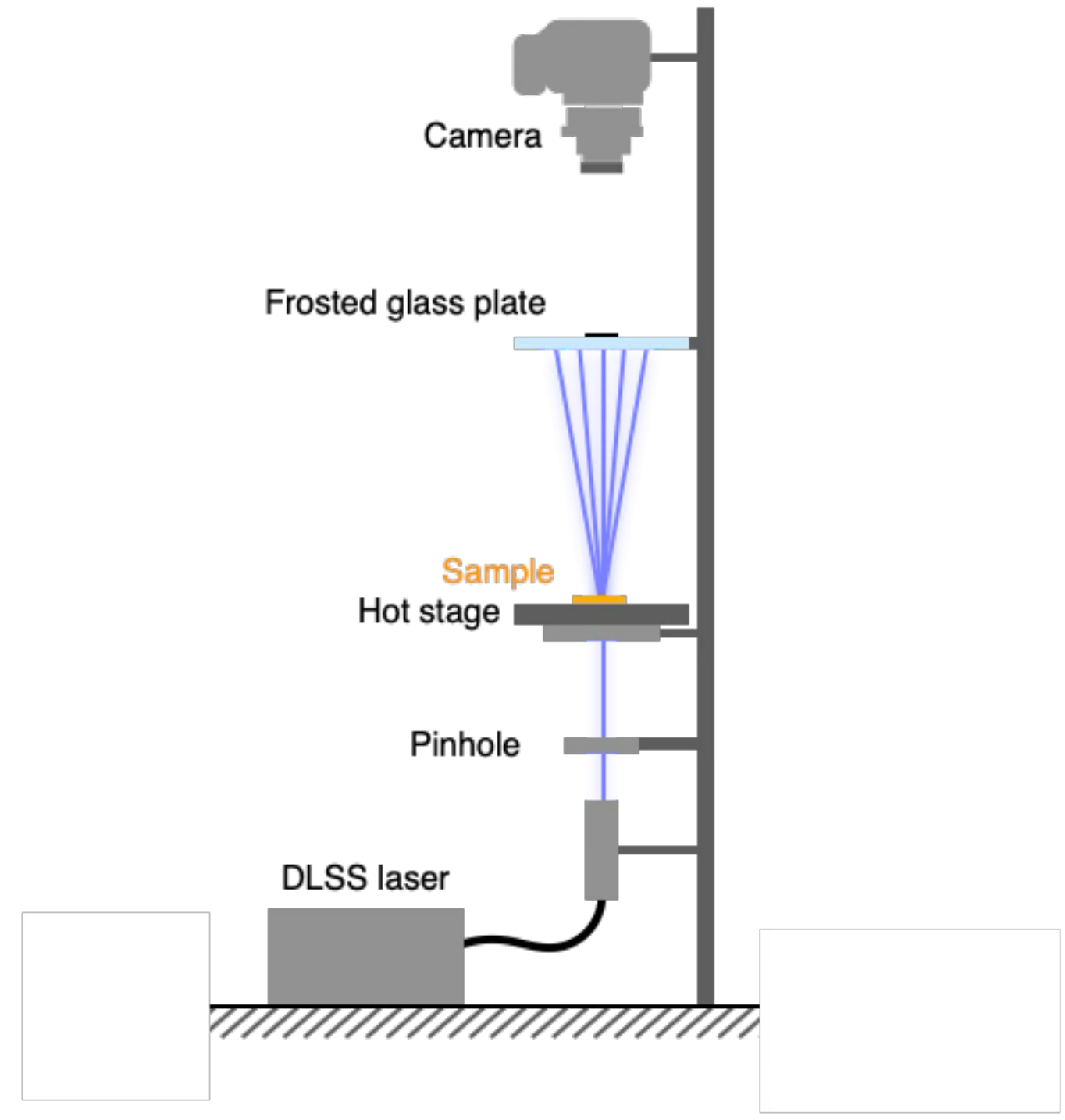

**Figure S34 |** Measurement setup for diffraction measurement.

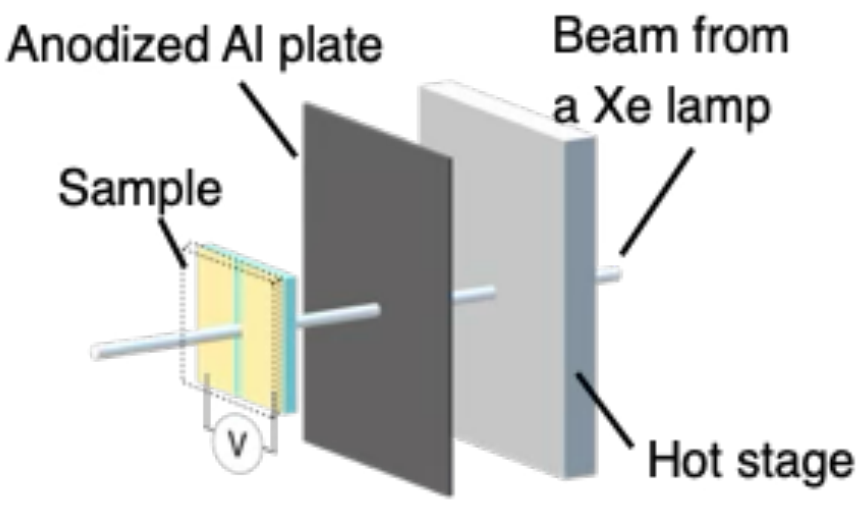

**Figure S35 |** A key geometry for CD spectra studies.

## Supporting Tables (Table S1, S2)

**Table S1 |** The molecular parameters of energy-minimized conformations calculated by MM2/DFT for **1**-*n* (n = 1–3).

| Entry | Vector X | Vector Y | Vector Z | $\mu$ (D) | $\beta$ (deg)† |
|---|---|---|---|---|---|
| **1**-*1* | 9.2896 | –3.1102 | –1.3125 | 9.88 | 19.97 |
| **1**-*2* | 9.4972 | –2.9639 | –1.3788 | 10.18 | 18.99 |
| **1**-*3* | 9.6458 | –2.8264 | –1.3728 | 10.24 | 18.04 |

† An angle between the permanent dipole moment ($\mu$) and long molecular axis.

**Table S2 |** Phase transition temperature (°C) and enthalpy changes (kJ $mol^{-1}$, in parenthesises) for **1**-*n* (n = 1–3).

| n | Process | K | | HEC | | $N_F$ | | N | | IL |
|---|---|---|---|---|---|---|---|---|---|---|
| 1 | 1C | • | 110.2<br>(15.4) | • | 122.5 | • | 172.3<br>(1.42) | • | 228.3<br>(1.67) | • |
| | 2H | • | 111.5<br>(23.3) | • | 122.7 | • | 164.4<br>(1.46) | • | 226.2<br>(1.67) | • |
| 2 | 1C | • | 106.0<br>(6.45) | • | 150.5 | • | 167.6<br>(2.17) | • | 220.7<br>(1.61) | • |
| | 2H | • | 109.5<br>(35.5) | • | 150.5 | • | 162.0<br>(2.19) | • | 219.2<br>(1.63) | • |
| 3 | 1C | • | 89.5<br>(36.1) | • | 146.9 | • | 150.8<br>(4.07) | • | 211.0<br>(2.42) | • |
| | 2H | • | 100.1<br>(37.6) | • | 146.9 | • | 147.9<br>(4.09) | • | 209.7<br>(2.45) | • |

Abbrev.: IL = isotropic liquid, N = nematic, $N_F$ = ferroelectric nematic, HEC = helielectric conical mesophase, K = crystal.

**Supporting Videos (Video S1–S4)**

**Video S1** | POM texture changes during phase transition from the N to HEC phases upon cooling. This video was taken under crossed polarizers in the antiparallel rubbed cell (thickness: 5 μm).

**Video S2** | POM texture changes during DC *E*-field sweep (0 V mm$^{-1}$[†] → +1.0 V mm$^{-1}$ → 0 V mm$^{-1}$ → −1.0 V mm$^{-1}$ → 0 V mm$^{-1}$). *E*-field steps: 0.1 V mm$^{-1}$. This video was taken under crossed polarizers in the antiparallel rubbed cell (thickness: 5 μm). The electrodes are located at both sides of the video recording area, but note that they are cut off. [†] Before recording, a pre-treatment of $E = -1.0$ to 0 V mm$^{-1}$ was taken place.

**Video S3** | Interchangeable diffracted patterns projected on the screen during DC *E*-field sweep (0 V mm$^{-1}$ → +1.0 V mm$^{-1}$ → 0 V mm$^{-1}$ → −1.0 V mm$^{-1}$ → 0 V mm$^{-1}$). *E*-field steps: 0.1 V mm$^{-1}$.

**Video S4** | POM texture changes during DC *E*-field sweep (0 V mm$^{-1}$ → +0.5 V mm$^{-1}$ → 0 V mm$^{-1}$ → −0.5 V mm$^{-1}$ → 0 V mm$^{-1}$). *E*-field steps: 0.01 V mm$^{-1}$. The electrodes are located at the top and bottom of the video recording area, but note that they are cut off.

**Supporting References**